\documentclass[journal abbreviation, manuscript]{copernicus}

\nolinenumbers

\usepackage{color}

\begin{document}

\title{A Practical Introduction to Regression-based Causal Inference in Meteorology (I): All confounders measured}

% \Author[affil]{given_name}{surname}

\Author[1,2][marzban@stat.washington.edu]{Caren}{Marzban} %% correspondence author
\Author[1]{Yikun}{Zhang}
\Author[3]{Nicholas}{Bond}
\Author[4]{Michael}{Richman}

\affil[1]{Department of Statistics, University of Washington, Seattle, Washington, 98195 USA}
\affil[2]{Applied Physics Laboratory, University of Washington, Seattle, Washington, 98195 USA}
\affil[3]{Climate Impacts Group, University of Washington, Seattle, Washington, 98195 USA}
\affil[4]{School of Meteorology, University of Oklahoma, Norman, Oklahoma, 73019 USA}

%% If an author is deceased, please add \deceased[$Deceased date if applicable$]{$Author number$} (e.g. \deceased[13 November 2015]{2}) at the end of the affiliations. The author number depends on the placement of the author in the author list, e.g. the third author has number 3.

%% If authors contributed equally, please add \equalcontrib{$Author numbers$} (e.g. \equalcontrib{1,3}) at the end of the affiliations. The author number depends on the placement of the author in the author list, e.g. the third author has number 3.

\runningtitle{Regression-based Causal Inference I}

\runningauthor{Marzban et al.}

\received{}
\pubdiscuss{} %% only important for two-stage journals
\revised{}
\accepted{}
\published{}

%% These dates will be inserted by Copernicus Publications during the typesetting process.

\firstpage{1}

\maketitle

\begin{abstract}
Whether a variable is the cause of another, or simply associated with it, is often an important scientific question. Causal Inference is the name associated with the body of techniques for addressing that question in a statistical setting. Although assessing causality is relatively straightforward in the presence of temporal information, outside that setting - the situation considered here - it is more difficult. The development of the field of causal inference has involved concepts from a wide range of topics, thereby limiting its adoption across some fields, including meteorology. However, at its core, the requisite knowledge involves little more than basic probability theory and regression, topics familiar to most meteorologists. By focusing on these core areas, this and a companion article provide a stepping stone for the meteorology community into the field of (non-temporal) causal inference.  Although some theoretical foundations are presented, the main goal is the application of a specific method, called matching, to a problem in meteorology. The data for the application are in the public domain, and R code is provided as well, forming an easy path for meteorology students and researchers to enter the field. 
\end{abstract}

\vspace{1.0cm}

\centerline{ \includegraphics[height=1.0in, width=1.5in]{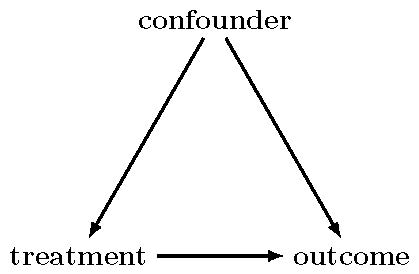} }

% \copyrightstatement{TEXT} %% This section is optional and can be used for copyright transfers.

\introduction  %% \introduction[modified heading if necessary]

It is well known that an association between two variables is necessary, but not sufficient, for
concluding that one variable causes the other. In much of the physical sciences, including the
atmospheric sciences, the application of statistics often involves associations, sufficient when
prediction is the main goal. But if the goal of a study is to gain diagnostic understanding of the
underlying causal structure, it is necessary to employ a body of knowledge generally referred to as
Causal Inference \citep{holland, imbens_rubin}.

The concept of causality has been the subject of philosophical exchange for ages; see Aristotle, Hume,
and Mill in \citep{holland}. However, the statistical origins of causal inference can be traced to
\citep{neyman}.  A key challenge in advancing an association between two variables to a causal
relationship lies in the potential presence of other variables - known as lurking variables - that
may be associated with both observed variables, thereby confounding the relationship.
For example, it is possible that an observed health benefit of exercise
is because those who exercise may be younger. In the jargon of statistics, exercise is
called a {\it treatment}, the individual receiving the treatment is called a {\it unit}, health is
an {\it outcome}, and age is a {\it confounder,} a special type of
lurking variable that causes both the treatment and the outcome.

A different type of lurking variable is called a mediator (Figure 1). In the context of the above
example, with exercise as the treatment, and health as the outcome, a mediator could be 
cardiovascular fitness. Note that cardiovascular fitness does not cause one to exercise, and so
it is not a confounder. Rather, it is more likely that exercise (the treatment) causes 
cardiovascular fitness, which in turn causes improved health. Methods for the causal analysis of 
mediators and confounders are different. The former is discussed in \citet{mediation}, while the
latter is the topic of the present work.  A more nuanced definition of a confounder is given in 
\citep{vanderWeele}. A glossary of some of the commonly used phrases in causal inference is 
provided in Appendix A.

When the treatment and outcome variables are available in the form of time series, it is relatively
simple to assess causality; two statistical tests commonly used in environmental sciences are the
test for Granger causality \citep{granger}, and the test for convergent cross mapping \citep{sugihara, 
tsonis}. However, in the absence of time series data - the situation studied in this and the sister 
article \citep{marzban_iv} - the most straightforward
method to ensure that confounding variables do not influence the relationship between treatment and
outcome is the implementation of a Randomized Controlled Experiment \citep{neyman, fisher}.
For example, to assess the causal effect of exercise on health, one begins with a random
sample of individuals representative of the population of interest. The sample is then randomly
divided into two groups, after which one group (the {\it treatment group}) is asked to exercise, and
the other group (the {\it control group}) is asked not to exercise. Any health difference between the
two groups is then likely to be caused by exercise because on average the two groups are similar
as a result of the initial randomization. As effective as such methods are in assessing causal
relations, they suffer from numerous shortcomings, including, for example, that the initial
randomization is often not possible on physical and/or ethical grounds. These shortcomings have led to
the development of methods for assessing causality in {\it observational data} wherein the
treatment is not randomly assigned to units, as is the case in the present work.

\begin{figure}[t]     % fig1
 \center
  \includegraphics[height=1.5in, width=3in]{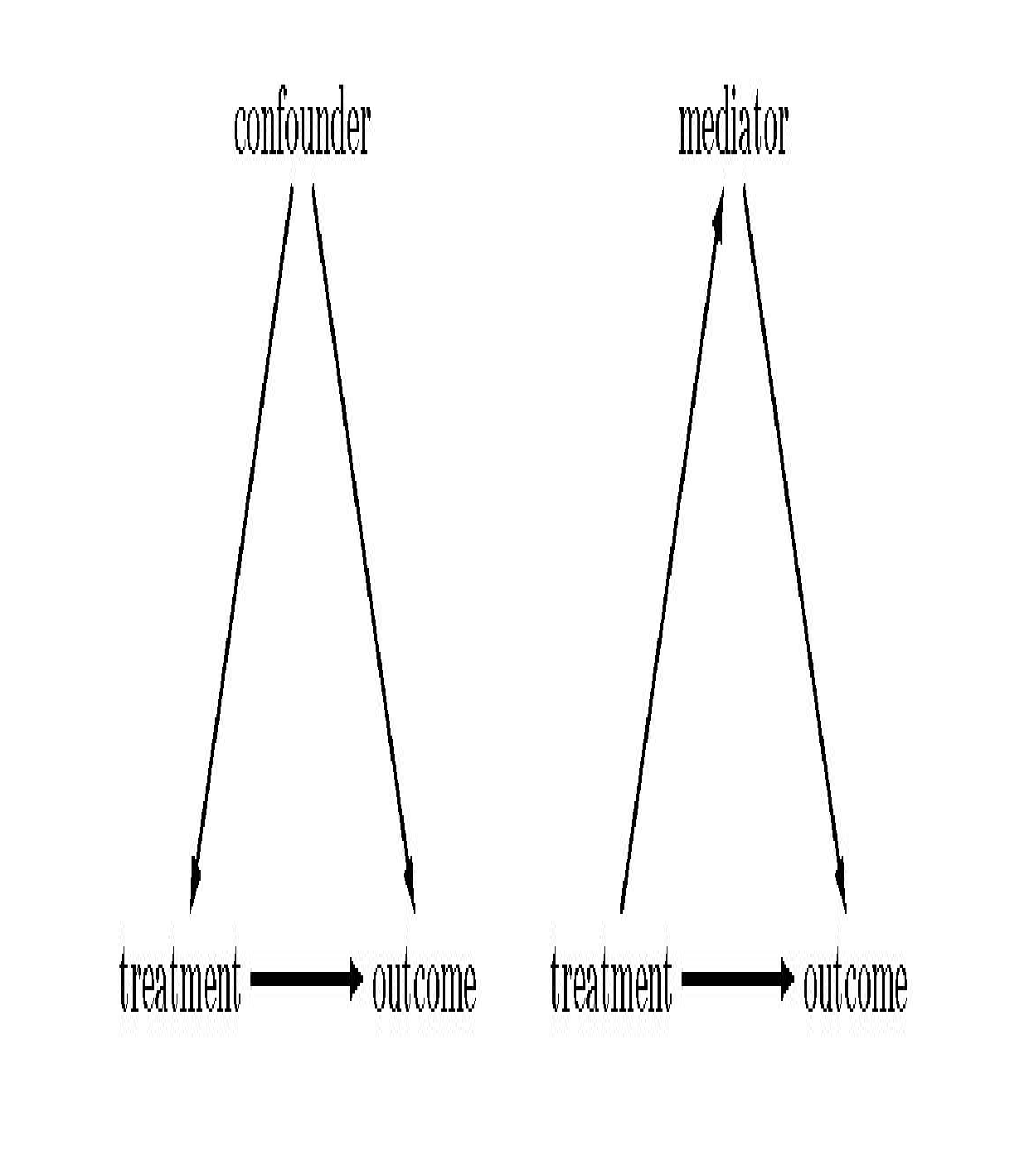}
\caption{Two examples of a lurking variable - a confounder (left), and a mediator (right).
Arrows indicate the direction of causality.}
\end{figure}

Some of the early pioneering work in causal inference is credited to \citet{cochran}, \citet{robins1986}
\citet{rubin_a, rubin_b}, and \citet{holland}; and a somewhat different framework is proposed by 
\citet{pearl2009, pearl2010}. 
For a more complete account of the history, see \citet{camps}. Much of the work has developed
in Economics and Public Health, fields wherein it is necessary to determine whether a 
treatment (also called an intervention)
has the desired causal effect. A classic example is the analysis of the so-called Lalonde data set,
wherein data from an observational study are employed to assess the effect of a job training program
on earnings \citep{lalonde}. Without the methods of causal inference, job training is found to have an
adverse effect on earnings. However, after accounting for confounders, the effect is found to be
positive \citep{dehejia, ding_book, imbens_xu, lalonde}. Such findings have important public policy
and public health implications. In another classic example, the effect of first-class seating on
the proportion of survivors on the Titanic falls from 35\% to 16\% after the confounding effect of
gender is taken into account \citep{mixtape}.

The more recent development of causal inference has spanned a wide range of fields, leading to
different frameworks. At the broadest level, they are referred to as the potential outcomes 
framework
and the graphical model framework. The two frameworks are somewhat distinct, with advantages and
disadvantages associated with both, and there have been attempts to unify them
\citep{richardson_robins, richardson_etal}.  For a more complete taxonomy,
see \citet{pearl2009, zeng_wang}.

The graphical model framework relies fundamentally on the concept of conditional independence, which
serves as the foundation for algorithms aimed at identifying causal relationships within data
\citep{zanga}. The algorithmic nature of this line of research and its reliance on
machine learning methods \citep{li_chu} set it apart from the
potential outcomes framework, which is generally more regression-based. The graphical model approach
has been liberally employed in climate science \citep{camps, ebert_deng, hannart2016, hannart2019, hirt, kretschmer, massmann, melkas, nowack, zeder}, but neither 
approach appears to have a clear presence in meteorology (at least in the non-time-series setting).

The heterogeneous nature of causal inference, developed across a wide range of fields, including 
Epidemiology, Economics, Public Health, Education, and Statistics, can be an obstacle to entry into 
the field. The
contemporary perspective of causal inference calls for a relatively deep understanding of concepts
ranging from statistics (e.g., inference in regression), probability theory (e.g., conditional
independence), graph theory (e.g., directed, acyclic graphs), and econometrics (e.g., instrumental
variables). Given that the meteorology community is generally familiar with regression, the present
and the accompanying paper \citep{marzban_iv} introduce two regression-based causal inference
methods, with the hope of serving as an avenue for meteorologists to enter the field of causal
inference. The examples considered here employ publicly available data, and R code is provided
 not only for generating all of the results, but also for further exploration. The application 
of causal inference methods based on regression
(i.e., a simple, robust, and time-tested method) to gridded data (arguably, one of the most
common forms of data in meteorology), is one of the novel features of both papers.

It is important to mention two caveats that have been put forth in the causal inference literature.
First, it has been argued that the causal structure underlying the variables (e.g., shown in Figure 1)
must be theoretically justifiable \citep{holland}. Said differently, there is nothing that prevents the
application of causal inference methods to variables that are in fact not causally related, thereby
supporting or contradicting hypotheses that have no causal underpinning at all. As such, it is 
important for the causal structure to be established {\it a priori,} e.g., through domain knowledge. 
Second, some authors have
argued that causal inference methods must not be applied to observational data if a randomized
experiment is not possible, at least in principle. \citet{cochran}, paraphrasing \citet{dorn}, states
\begin{center}
\emph{The planner of an observational study should always ask himself the question, ``How would the 
study be conducted if it were possible to do it by controlled experimentation?''}
\end{center}
The phrase ``no causation
without intervention,'' aims to capture that sentiment \citep{holland, rubin_a, rubin_b, rubin2007}.
In that quote, although the word ``intervention" implies a direct and active manipulation of the 
treatment assignments, there are some controversies surrounding that interpretation; see the exchange
between A. Gelman and J. Pearl, where the latter argues that ``The mantra 'No Causation without 
Manipulation' is a relic of a bygone era.'' \citep{gelman_blog}; see the Closing Remarks section 
here for more on the notion of intervention. 

The structure of this paper is as follows: The method section begins by presenting the
potential outcomes framework for causal inference, in general, then focuses on the method of matching,
presents the concept of balance, which is central to the matching method, and ends with the
presentation of two regression models for estimating the average causal effect.\footnote{Although the
matching method is often considered to belong to the domain of machine learning \citep{li_chu},
no machine learning is required in this article.}
The data section provides the theoretical and empirical justification for the meteorological
variables selected for the demonstration. Details of the matching method and the assessment of
balance are presented in the Results section, followed by a summary of the conclusions,
and proposals for generalization and future work. Some of the more subtle caveats of the approach
are discussed in the Closing Remarks section.

\section{Method}

\subsection{The Theory}

The methods of causal inference vary by whether the three random variables - treatment ($A$),
confounding ($X$), and outcome ($Y$) - are binary or continuous. Here the treatment is assumed to be
binary. A sister article by the authors of this work can also deal with situations where the treatment
is continuous \citep{marzban_iv}; another important distinguishing characteristic of the two papers
is as follows: in the present work the confounders are assumed to be measured, while in the
accompanying article they are unmeasured. The confounding and the outcome variables are assumed to
be continuous. Moreover, there may be more than one confounding variable present, in which case
$X$, as a vector, will denote all of them.

As mentioned in the Introduction, there are several frameworks for defining causality. Here, the
potential outcomes framework is considered, where one assumes that for the $i^{th}$ unit there exist two
possible states, denoted 0 and 1, corresponding to the control and treatment groups, respectively.
The corresponding {\bf potential} outcomes are denoted $Y_i(0), Y_i(1)$, and the {\bf observed}
outcome is denoted $Y_i$. It is important to emphasize that only one of the potential outcomes is
observed for each unit. The word counterfactual is used to refer to the unobserved potential outcome.

Although a clear statement of the assumptions underlying a method is necessary in any endeavor, it is
especially important in causal inference. Indeed, a great deal of the work is dedicated to determining
the conditions under which the identification of causality from observed data is possible at all
\citep{ding_book, hernan_whatif, morgan_book, gelman_book}, after which
those conditions are either assumed to hold, or modeling steps are taken to assure that the data do not
contradict the conditions.  Two of the more prominent conditions are
\begin{eqnarray}
Y_i = \left\{ \begin{array}{lll}
 Y_i(0) & \mbox{if} & A_i=0  \\
 Y_i(1) & \mbox{if} & A_i=1 \; ,
                \end{array}  \right. \\
 (Y_i(1), Y_i(0)) \perp A_i \; | \; X_i \;.
\end{eqnarray}
where the symbol $\perp$ denotes independence.
The first condition is called the consistency assumption, essentially providing a link between the
potential and the observed world.\footnote{There is some controversy regarding the nature of the
statement in Eq. 1. Some authors (e.g., \citet{cole_frangakis} and \citet{pearl_consistency}) consider 
it an important assumption, while others (e.g., \citet{ding_book}) consider it the definition of the 
observed outcome. Also, see the Closing Remarks section for a different use of the name ``consistency.''}

The second condition - called by a variety of names,
including conditional independence, ignorable treatment assignment, and No Unmeasured Confounding
Assumption (NUCA) - plays an
important role in allowing for the estimation of a causal effect with observational (i.e.,
non-randomized) data. It requires that the potential outcomes are independent of the treatment, given
the confounders $X$. This assumption may seem to contradict the expectation that a treatment and the
outcome ought to be at least associated. The resolution to this apparent contradiction is in the
realization
that it is the observed outcome, not the potential outcome, that is expected to depend on the treatment.
As suggested by its name, NUCA requires that all confounders have been identified and measured.
A violation of this assumption implies the existence of confounding, which in turn leads to a biased
estimate of the average causal effect.

To demonstrate the importance of the consistency assumption, consider that a quantity of
interest is the Average Treatment Effect (ATE), defined as $E[Y_i(1) - Y_i(0)]$.
A simplifying assumption is that the ATE is uniform across all units, in which case, and
henceforth, the subscript $i$ will be suppressed. The expectation operator $E[.]$ is an
average over a distribution/population, generally estimated with a sample average and then 
supplemented with a confidence interval.\footnote{The notation adopted in this paper is somewhat
oversimplified in that it blurs the distinction between a random variable, its realizations, and 
their specification to the $i^{th}$ unit. For a rigorous treatment, see pages 33-35, and especially
footnotes 5 and 6 in \citet{morgan_book}, specifically ``Accordingly, we will denote individual-level
potential outcomes as values such as $y^1_i$ and $y^0_i$, regarding these as realizations of
population-level random variables $Y^1$ and $Y^0$ while recognizing, at least implicitly, that they
could also be regarded as realizations of individual-specific random variables $Y^1_i$ and $Y^0_i$.''}

The ATE cannot be estimated because only one of the two potential outcomes is observed.
This obstacle is called ``The fundamental problem of causal inference.'' By contrast, what can
be estimated is $E[Y | A = 1] - E[Y | A = 0]$. It can be shown that this simple difference in means
can be decomposed as follows (\citep{mixtape}; also see Appendix B):
\begin{eqnarray}
E[Y(1) | A = 1] - E[Y(0) | A = 0 ] &=& ATE \\ \nonumber
                                   &+& ( E[Y(0)|A=1] - E[Y(0)|A=0])\\ \nonumber
                                   &+& (1-\pi)(ATT-ATC),
\end{eqnarray}
% 2nd term = selection bias , 3rd term = heterogeneous treatment effect bias
where $\pi$ is the probability that a unit will receive the treatment, and ATT and ATC are conditional
average treatment effects for the treatment group ($A=1$) and the control group ($A=0$), respectively.
Note that by the consistency assumption the left-hand side of Eq. (3) can be written as
$E[Y \; | \; A = 1] - E[Y \; | \; A = 0]$, a simple difference in means that can be estimated from
observed data.  This decomposition is important because it highlights the difference between
the average effect that can be estimated from the observed data (i.e., the left-hand side), and the
average effect of interest (i.e., ATE), which cannot be estimated directly. However, if it can be
arranged for the second and third terms on the right-hand side of the equation to be zero, then ATE can be
estimated with the simple difference in means.

Indeed, it can be shown that the last two terms in Eq. (3) are zero if $(Y(0), Y(1))$ are independent
of $A$ \citep{mixtape}. This
independence assumption is the reason in a randomized experiment the observed difference in means
can be taken as a measure of a causal effect. Randomization of the treatment across the units assures
that the treatment and control groups are exchangeable, leading to the independence of the treatment
and the potential outcomes.

In situations where this randomization is absent (e.g., in observational data), the second assumption
in Eq. (2) plays an important role. Specifically, note that if there exist confounders $X$ satisfying
Eq. (2), then
\begin{eqnarray}
 E[Y(0) \; | \; A=0, X] = E[Y(0) \; | \; X], \\
 E[Y(1) \; | \; A=1, X] = E[Y(1) \; | \; X],
\end{eqnarray}
in which case
\begin{eqnarray}
E[ Y(1) - Y(0) \; | \; X ] &=& E[ Y(1) \; | \; A = 1, X] - E[Y(0) \; | \; A = 0, X ]\\ \nonumber
                           &=& E[Y \; | \; A = 1, X] - E[Y \; | \; A = 0, X],
\end{eqnarray}
where the second equality follows from the consistency assumption Eq. (1).  This time, note that the
right-hand side of Eq. (6) can be estimated from data, and the average of the left-hand side
over the confounders yields ATE. In short, if the consistency and conditional independence
assumptions are satisfied, then ATE can be estimated by averaging the simple difference in means over
the possible values of $X$, a method known as g-computation \citep{robins}, i.e.,
\begin{equation}
ATE = E_X \left[ E [Y | A = 1, X] - E [Y | A = 0, X] \right].
\end{equation}

The goal of estimating the ATE becomes possible under the consistency and the conditional independence 
assumptions in Eqs. (1) and (2). Equations (4) and (5) express the consequence of this assumption: 
within levels of the confounders $X$, the mean potential outcome is the same for units that did and 
did not receive the treatment. It is important to note, however, that these conditions involve potential outcomes, only one of which is observed for each unit, and therefore cannot be tested directly.

Because the conditional independence assumption itself is untestable, practical implementations of 
causal inference often focus on design-based diagnostics that are observable. In particular, if 
adjustment for $X$ is to remove confounding, then the treated and control groups should be made 
comparable with respect to the observed confounders. In a randomized experiment, this comparability 
is achieved, in expectation, by random assignment: the treatment indicator $A$ is independent of the 
confounders $X$, so that $\Pr(X\mid A=0)=\Pr(X\mid A=1)$.  In that case, the treated and control 
groups are said to be balanced with respect to $X$.

In an observational study, however, $A$ and $X$ are generally not independent, and so 
$\Pr(X\mid A=0)$ and $\Pr(X\mid A=1)$ may differ substantially. Even if $X$ and $A$ are 
not assured to be independent, it is possible that $X$ and $A$ are independent, given some function 
of $X$ (denoted $A \perp X \; | \; f(X)$), in which case the function $f(X)$ is said to be a 
{\it balancing score,} one of the building blocks of causal inference \citep{dawid}. As described 
further in the next two subsections, the purpose of matching or weighting is to construct a matched 
sample in which the distribution of the measured confounders is more similar across treatment groups. 
This balance is not implied by Eqs. (4) and (5); rather, it is an observable diagnostic used to assess 
whether the matching or weighting procedure has succeeded in making the treated and control groups 
comparable with respect to the measured confounders.

An explanation of $A \perp X \; | \;f(X)$ is in order. Superficially, it may seem nonsensical because
if $f(X)$ is given, then so is $X$, in which case the independence of $X$ and $A$ simply makes no sense.
At this point it is important to note that in general $X$ is a random {\it vector}. In other
words, $X$ denotes not a single random variable but a multivariate quantity, consisting of multiple
confounders $X_1, X_2, \ldots$. In this multivariate situation, $X$ is a vector, but $f(X)$ is still
a scalar. Therefore, knowledge of the single number $f(X)$ does not uniquely identify the vector
$X$. As such, it makes perfect sense to write expressions like $A \perp X \; | \; f(X)$.

In principle, one can search for functions that are balancing scores, and then proceed with the
computation of ATE, using the g-computation formula described above. But in situations where there 
are a large number of
confounders, the ``curse of dimensionality'' can lead to imprecise estimates.  A remarkable theorem
due to \citet{rosenbaum_rubin} circumvents this problem. To that end, one defines a quantity called
the {\it propensity score} (PS), defined as $pr(A=1 \; | \; X)$ and denoted $p(X)$. It can be shown that
$p(X) = pr(A=1 \; | \; X_1, X_2, \ldots)$, is a balancing score. In short, $A \perp X \; | \; p(X)$.
The advantage of the PS is that it is a single, scalar function which can be computed even in the
presence of multiple confounders. This construction of the PS is particularly attractive because it
can be estimated by performing regression on the treatment (as a binary response) and all confounders
as covariates/predictors.\footnote{Generalizations of the propensity score to non-binary and 
continuous treatment variables can be found in \citep{hirano_imbens} and \citep{kosuke_vandyke}.}

In summary, for a given PS - a quantity that can be estimated from data on the treatment
variable and the confounders - the treatment and the confounders are independent. Therefore,
if one were to group/match the control and treatment units such that they all have the same (or similar)
PS, then within each group the confounders are independent of the treatment, and thus, cannot
act as lurking variables.

\subsection{Matching}

A large number of matching methods have been proposed. This section provides an intuitive discussion
of the basic idea. More details can be found in \citep{rubin_a, rubin_b} and \citep{stuart}.

To highlight the basic steps of matching, consider the example in the Introduction, where the two
treatment groups are those who exercise and those who do not.  Consider one of the units in the
exercise group. If this person is young, then match that person with a young person in the non-exercise 
group; otherwise, match that person with an older person in the non-exercise group.
This matching leads to pairs of individuals both of whom have the same age, but only
one exercises. As such, any difference in health between the pair cannot be attributed to
age. This type of matching is not limited to pairs of individuals and can be generalized to
include multiple units. Moreover, multiple lurking variables can be used to perform the matching.
Although matching offers a principled approach to eliminating the effects of the confounding variables,
it could still suffer from the same ``curse of dimensionality'' that afflicts the method of 
controlling multiple confounding variables.  The ``magic" of matching is that it is not necessary 
to match the cases based on all of the confounding variables, but only based on one 
variable - the PS \citep{heinrich, abadie_imbens}. Intuitively, the reason it is sufficient to
match with PS only is that the PS is a balancing score, i.e., conditional on the PS, the distribution 
of all measured confounders is the same in the treatment and control groups. Therefore, units with 
similar PS can be compared as if treatment assignment were randomized with respect to 
those confounders \citep{rosenbaum_rubin}.

When the confounder is a continuous variable, the matching method can be explained in terms of
conditional histograms.  By definition, a confounder is associated with the treatment variable,
and therefore, the histogram of the confounder in the control group is different from that in the
treatment group. For example, the probability of a specific value of the confounder may be low for
the control group but high for the treatment group. The idea of matching is to create a data set,
called a {\it pseudo-population}, in which that specific value of the confounder appears with equal
or at least comparable probability in both groups. This can be arranged by re-sampling (with 
replacement) the observational
data such that each unit is sampled with a frequency weighted by the inverse of the
probability \citep{rosenbaum_rubin}. This procedure will lead to a simulated data
set in which the confounder's histogram in the treatment group is comparable to that in the control
group. Said differently, in the generated pseudo-population,
on average the two groups are balanced with respect to the confounder. Then,
any difference in the outcome across the treatment and control groups must be due to, or caused by, 
the treatment.  This perspective of matching is further elaborated in the Balance section, below.

There exist a large number of variations on the basic notion of matching, but the final ``output'' in
most of them is the aforementioned weights, one per unit. As explained previously, it can be shown that
it is not necessary to perform the matching for all confounders; it is sufficient to generate the
weights by matching the PS only. The formulas for generating the weights depend on
whether one is interested in estimating ATE, ATT, or ATC, because they are all computed from the
pseudo-population. Although the computation of these matching weights is not elaborated upon here,
it is essential to all aspects of matching methods.  In short, an important step in matching
methods is to assure that the treatment and control group in the pseudo-population are balanced 
in terms of the confounders. Also note that the matching process involves only the
confounders and the treatment variable; the outcome variable does not enter the matching procedure
at all.

\subsection{Balance}

Numerous methods have been proposed for testing the balance of the confounders
\citep{greifer_stuart, greifer}, i.e.,
the extent to which the following equation is satisfied: $pr(X \; | \; A=0) = pr(X \; | \; A=1)$.
As explained above, one method aims to generate a pseudo-population in which the conditional histogram
of confounding variables in the treatment group is similar to that
in the control group. A natural device for the comparison of two histograms is the two-sample Q-Q plot,
wherein one plots the quantiles of a confounder when $A=1$ versus those of the confounder when $A=0$.
One then compares the Q-Q plot for a given confounder before matching with that after matching; a
near-diagonal Q-Q plot is evidence of a good match. By virtue of being a graphical
tool, Q-Q plots have diagnostic utility in terms of suggesting steps that may improve the balance.

If the number of confounders is large, the visual means of assessing balance become unwieldy. Instead,
one simply examines the conditional mean of each confounder for the two groups. A common summary
measure is the Standardized Mean Difference (SMD), defined as the difference of the conditional
means, divided by the pooled standard deviation.\footnote{This standardization is appropriate only
when estimating the ATE; estimation of ATT and ATC requires a different standardization \citep{greifer}.}
Then, one compares SMD before matching with that after matching, and a small SMD is indicative of good
matching.

At this point, it is important to address a computational issue.  The SMD involves the difference
of two conditional means of a confounder. Before matching, each conditional mean is a simple average
of a confounder in the treatment and control groups. After matching, the conditional means are
computed as weighted means, with the weights derived from the matching process (and the
treatment model, described next). All of these calculations are
straightforward because they involve the original data and the weights assigned to each observation.
Even the computation of the Q-Q plot before matching is straightforward because again it involves the
observed confounder in the two treatment groups. However, a Q-Q plot after matching, requires
``observations'' of a confounder after matching; to that end, the aforementioned pseudo-population
is employed.

\subsection{The Treatment Model, Matching, and The Outcome Model}

The overall process of matching involves three steps: 1) Estimating the PS, 2) Matching with PS,
and 3) estimation of ATE.

For the first step, since PS is defined as the probability of $A=1$, given the confounders, it is
natural to estimate it
from a regression model with the treatment as the response, and the confounders as covariates/predictors.
Given that the treatment $A$ is binary, the link function may be a logit or probit function. Here, and
without any particular preference, the latter is used, i.e.,
\begin{equation}
pr(A=1 \; | \; X) = \Phi(\gamma_0 + \gamma_1 X_1 + \gamma_2 X_2 + \cdots),
\end{equation}
where $\Phi(u)$ denotes the normal cumulative distribution function, and the parameters $\gamma_0,
\gamma_1, \gamma_2, \cdots$ are estimated via the maximum-likelihood criterion. Since the model
estimates the probability of receiving the treatment, given the confounders, this regression model is
called the {\it treatment model}; it corresponds to the line connecting the confounder to the treatment
in Figure 1.  The prediction of the treatment model for every unit in the data estimates the PS
for each unit.

The resulting PS values are then used to perform the matching step. For the many options available 
in this step, see the R code provided in the supplementary material. For example, in what is called 
``full matching'' (used here), one partitions the sample into a number of matched strata, where 
each stratum contains at least one unit from the treatment group and at least one control unit.
The matching is performed so as to minimize the within-stratum difference in the PS.

In order to motivate how matching leads to weights, it helps to consider a sampling problem. 
An estimator of a mean is biased if the units have unequal probability of being selected. However, 
an unbiased estimator can be constructed by simply computing a 
weighted mean, with the weights accounting for the unequal probabilities. One well-known estimator is 
the Hansen-Hurwitz estimator \citep{hansen-hurwitz}, where the weights are simply the inverse of the 
probabilities. Similarly, it can be shown that an unbiased estimator of ATE can be constructed 
\citep{stuart} by setting the weight for the $i^{th}$ unit to 
$\frac{A_i}{\hat{ps_i}} + \frac{1-A_i}{1-\hat{ps_i}}$, where $\hat{ps_i}$ is the
estimated PS for the $i^{th}$ unit, obtained from the treatment model in Eq. (8). This is how matching 
by the PS leads to a set of weights, one for each unit.\footnote{In the R function for matching, i.e., 
matchit(), in each stratum, a new PS is computed, called ''stratum propensity score,'' as the 
proportion of units in each stratum that are in the treatment group. Then, all units in each stratum 
are assigned the within-stratum PS.}

Armed with the matching weights the last step is the development of the so-called {\it outcome model}:
\begin{equation}
Y = \alpha_0 + \beta_0 A + \beta_1 X_1 + \beta_2 X_2 + \cdots + \alpha_1 A X_1 + \alpha_2 A X_2 +
\cdots ,
\end{equation}
where the regression coefficients are estimated via a weighted least-squares criterion, with the
weights provided by the matching weights.  An important facet of the outcome model is the presence
of treatment-confounder interactions, shown to be necessary if the resulting predictions are to
adequately estimate the potential outcomes \citep{greifer_stuart, greifer}.
With these parameter estimates, ATE can be estimated using the g-computation method described in the
Theory subsection above. This regression-based approach also allows
for analytic expressions for the confidence interval for ATE.

Equation (9) may appear to be a (non-causal) multiple regression model wherein one has adjusted
for the confounders. This similarity raises the question of how matching is different from adjusting.
In this specific example, the difference is in the presence of the weights employed in the weighted
least-squares criterion. Without these weights, the outcome model would be the traditional adjusted 
model considered in non-causal-inference analyses. More generally, the matching and adjusting methods
can be compared in a variety of ways.  At the simplest level, one can contrast them by noting that
adjusting amounts to ignoring the line connecting the confounder to the treatment variable in Figure 1.
More broadly, \citet{ho_etal} argue that
matching can be viewed as a nonparametric method for adjusting. A mathematical analysis of the
relationship between the two is presented in \citep{chatto}, and a
less formal comparison of their pros and cons is discussed by \citet{noah}. Another perspective
through which matching and adjusting can be compared is model misspecification; it is well-known
that if either the treatment model or the outcome model is incorrect, then the estimate of ATE
considered here may be biased \citep{gelman_book}. The impact of model misspecification on bias
is studied by
\citep{vansteelandt} who also argue that all methods of causal inference, including matching,
are designed to minimize this bias. In other words, matching alleviates the impact of bias on the
estimation of ATE in situations where the treatment model and/or the outcome model may be wrong.
Given that the matching and adjusting approaches involve different assumptions about the underlying
relationships, both estimates of ATE are presented here.

\section{Data}

The data employed for this presentation are from the North American Regional Reanalysis (NARR)
archive \citep{mesinger}, consisting of 418 variables computed on a $349 \times 277$ 
(longitude $\times$ latitude) grid.
Partly for reducing computational burden, and partly for increasing the prominence of land in the
region of analysis, 120 grid points are excluded from the western side of the domain, and 80 grid
points are excluded from the other three sides of the domain, leaving a domain of $149 \times 117$ 
grid points for analysis. Figure 4 shows the resulting domain.

To avoid temporal dependencies, data from a single January are analyzed - specifically, the 
monthly-averaged data for January 2001. Although this choice of the date is mostly random, 
preference is given to dates with fewer missing data. For instance, in several other dates, some of
the confounders are missing entirely. 
January 2001 was largely unremarkable in that it had no major ENSO activity, occurring at
the tail end of a La Ni\~na with Ocean Ni\~no Index values close to neutral, removing any
major effects of ENSO on the confounder, treatment, and outcome. Further, the patterns of
geopotential, temperature and outgoing longwave radiation (a measure of cloudiness) are
similar to the 1991-2020 30-year climatology. Thus, January 2001 is a representative
sample of the 30-year climatology.

Consequently, the experimental unit for the study is the 
grid point, and all correlations between variables are strictly spatial in nature. This spatial
nature of the correlations is important in the theoretical considerations necessary for
justifying the causal structure of the variables examined. Here, the variables examined are as 
follows, with theoretical justification provided in the next subsection:
\begin{itemize}
\item Treatment variable $(A)$ = downward shortwave radiation flux $(W/m^2)$, referred to as 
``radiation,'' when appropriate.
\item Outcome variable $(Y)$ = surface potential temperature $(^\circ C)$,\footnote{The word 
``potential'' in potential temperature must not be confused with the use of that word in reference to 
potential outcomes.} referred to as ``temperature,'' when appropriate.
\item Confounding variables $(X)$ = geopotential height $(gpm)$, referred to as ``height,'' when appropriate.
\end{itemize}

The NARR archive contains data on geopotential height at 29 pressure levels ($100 hPa,\; 125 hPa, \ldots ,
300 hPa,\; 350 hPa, \ldots,$\\
$700 hPa, 725 hPa, \ldots, 1000 hPa$), but only two levels are used in the 
analysis, specifically the $3^{rd}$ and the $24^{th}$, respectively corresponding to $150hPa$ and 
$875hPa$, and denoted $X_3, X_{24}$. These variables are selected based on theoretical and statistical 
considerations discussed in the following subsections.
The geopotential heights of these pressure levels are below surface elevations in regions of
higher terrain such as the Rocky Mountains; in those regions NARR essentially extrapolates their
values to a level higher than the actual pressure, which can lead to inaccuracies in the physical
interpretations using potential temperature. In general, their distributions tend to resemble,
but not strictly correspond with sea-level pressure, which of course is also an extrapolated variable.

\subsection{Theoretical Justification}

Theoretical justification for this choice of variables is as follows: These variables are known to 
have physical linkages between them. In particular, the downward shortwave radiation flux represents 
an important factor in the energy budget for the near-surface layer of the atmosphere. Greater 
solar insolation means greater heating of the ground, and ultimately warmer surface air temperatures, 
all other terms in the energy budget remaining unchanged. Here, rather than the surface air 
temperature itself, the surface potential temperature is used, with the latter representing the 
temperature a parcel of air would attain if brought adiabatically to the standard pressure level of 
$1000 mb$. In other words, because the potential temperature relates to the heat energy content, it 
can account for variations in surface temperature simply related to elevation and hence surface 
pressure.

During the month of January in the Northern Hemisphere there is the tendency for substantially lower 
values of mean insolation at higher latitudes than in the sub-tropics due to the shorter day lengths 
and lower sun angles. Of course, the downward shortwave radiation flux is by no means the only, or 
even necessarily the dominant mechanism controlling surface potential temperatures. Complicated and 
interacting physical processes, featuring the conservation of heat, momentum and angular momentum 
modified by diabatic and viscous effects, ultimately result in significant net transfer of heat 
poleward. Higher air temperatures imply less dense air, and from hydrostatic considerations, greater 
geopotential heights aloft. Relative to a static system, the transports of heat by the atmospheric and 
oceanic circulations thereby serve to raise geopotential heights aloft at higher latitudes and reduce 
heights at lower latitudes overall, particularly from the mid-tropospheric to stratospheric levels 
considered in the analysis that follows. It bears emphasizing that these effects are by no means 
zonally uniform, i.e., independent of longitude. The resultant patterns are dynamic rather than 
static, influenced by both terrain and surface characteristics (notably land versus ocean effects), 
and in turn correspond with the winds, precipitation, clouds and other phenomena impacting surface 
potential temperatures. And obviously, clouds modulate the downward shortwave radiation fluxes. 
In short, the latitudinal gradient of mean downward shortwave radiation at the top of the atmosphere 
sets the stage, but other factors play key roles in the distribution of surface potential temperatures.

A strong case can be made that geopotential heights at the pressure levels considered here are 
confounders rather than mediators, as defined by the direction of causality illustrated in Figure 1. 
From a pressure/mass perspective, roughly 10\% of the atmosphere consists of the atmospheric boundary 
layer (ABL) with a top typically on the order of $1000 m/100 hPa$ above the surface. The $875 hPa$ 
reference pressure level used here is therefore generally near or above the top of the ABL. Surface 
potential temperatures are directly related to vertical profiles in temperature and hence density due 
to mixing within the ABL, but this influence is largely restricted to the ABL. Heating (cooling) of 
surface temperatures leads to vertical expansion (contraction) of the atmospheric column within the 
ABL and higher (lower) geopotential heights at a pressure level near the top of the ABL, assuming 
negligible net horizontal transports of mass. But because the $875 hPa$ pressure level is underneath 
roughly 85\% of the atmosphere, the distributions of temperature/density aloft largely determine the 
geopotential height distributions at that pressure level, especially on monthly and longer time scales. 
As part of the present study, the distributions of monthly mean geopotential height at $1000 hPa$ 
were compared with their counterparts at $875 hPa$; the patterns at the two pressure levels were 
found to closely resemble one another with a simple offset, implying that ABL temperatures and hence 
surface potential temperatures had quite minor influences on the geopotential height at the 
$875 hPa$ pressure level. 

Regarding the distribution of geopotential heights at the $150 hPa$ pressure level, as mentioned
above, the geopotential heights at $150 hPa$ 
relative to the surface pressure are in hydrostatic balance with the distributions of 
vertically-integrated profiles of density, largely in association with temperature. ABL and surface 
temperatures also have minimal effects on the density profiles and ultimately $150 hPa$ geopotential 
height distributions which instead reflect zonal mean properties, i.e., temperatures and meridional 
circulations, subject to longitudinal variations. As shown in Figure 2 below, the distributions of 
geopotential height at $150 hPa$ and $875 hPa$ resemble rather than match one another, due to 
spatial variations in temperature/density between the two pressure levels.

To be confounders, the geopotential height distributions at the $875 hPa$ and $150 hPa$ pressure levels 
must have impacts on surface potential temperatures. First considering the $875 hPa$ level, 
spatial variations in geopotential height result in lower-tropospheric advection with impacts on 
near-surface temperatures, notably, but not exclusively, cold-air outbreaks from higher latitudes and 
warm surges from lower latitudes. While such events often occur on time scales of a few days, their 
net signatures are apparent on monthly and longer time scales. Moreover, the rising motion often 
associated with lower geopotential heights at lower elevations (including the $875 hPa$ pressure level) 
is instrumental in not only promoting the condensation of water vapor leading to clouds and hence the 
confounding effect of reduced shortwave radiation, but also precipitation, which in turn tends to lead 
to near-surface cooling due to evaporation and sublimation of hydrometeors. Geopotential heights at the 
$150 hPa$ level are related to large-scale circulations (e.g., the Ferrel and Polar cells) with direct 
impacts on temperatures aloft, and vertical motions that are primarily manifested in the middle and 
upper troposphere, and direct impacts on air temperatures aloft. Important consequences include the 
tendency for lesser (greater) surface heating due to the downward longwave radiation in regions of 
lower (higher) geopotential heights. Similarly, greater (lesser) extratropical cyclone development, 
with attendant lower tropospheric temperature advection and diabatic effects, tends to occur in 
regions of overall upward (downward) motion aloft. In short, it is expected {\it a priori} that 
spatial variations in geopotential height levels are much more a cause than an effect of 
corresponding distributions in surface potential temperature. 

Although in the present situation, as argued above, it is difficult to justify that both the downward 
shortwave radiation and surface potential temperature are primary causes of geopotential height, 
it is worth mentioning that if that were true, geopotential height would be a variable with arrows 
pointing to it in Figure 1. In causal inference jargon such a variable is called a collider (as opposed 
to a confounder). The distinction is important because it can be shown that conditioning on 
(or adjusting for) a confounder reduces bias, but conditioning on a collider increases 
bias \citep{pearl2009}.

\subsection{Statistical Justification}

The choice of geopotential height as a confounding variable is theoretically justified above, but 
there are also statistical considerations that must be made. For example, one may be tempted to
include all 29 pressure levels in the analysis. However, the corresponding 29 variables are highly
intercorrelated, and therefore, their inclusion in the statistical models can lead to overfitting
and imprecise estimation due to multicollinearity. The left panel in Figure 2 shows the correlations 
between all pairs of the 29 confounders and confirms their highly intercorrelated nature. Evidently, 
only the top few pressure levels have a correlation with other levels that are not near 1. As such, 
it is inappropriate to use all 29 confounders.

An eigendecomposition of the correlation matrix shown in Figure 2, shows that the cumulative 
percentage of total variance explained by the first two eigenvectors is 98.1\% and 99.5\%.
It may be tempting to use the two eigenvectors directly, but it is unclear how the projection of the 
data onto the principal components affects the causal structure of the variables involved. Instead, 
only two relatively uncorrelated confounders may be selected for the analysis.

\begin{figure}[t]   % fig2
\center
\includegraphics[height=4in,width=8in]{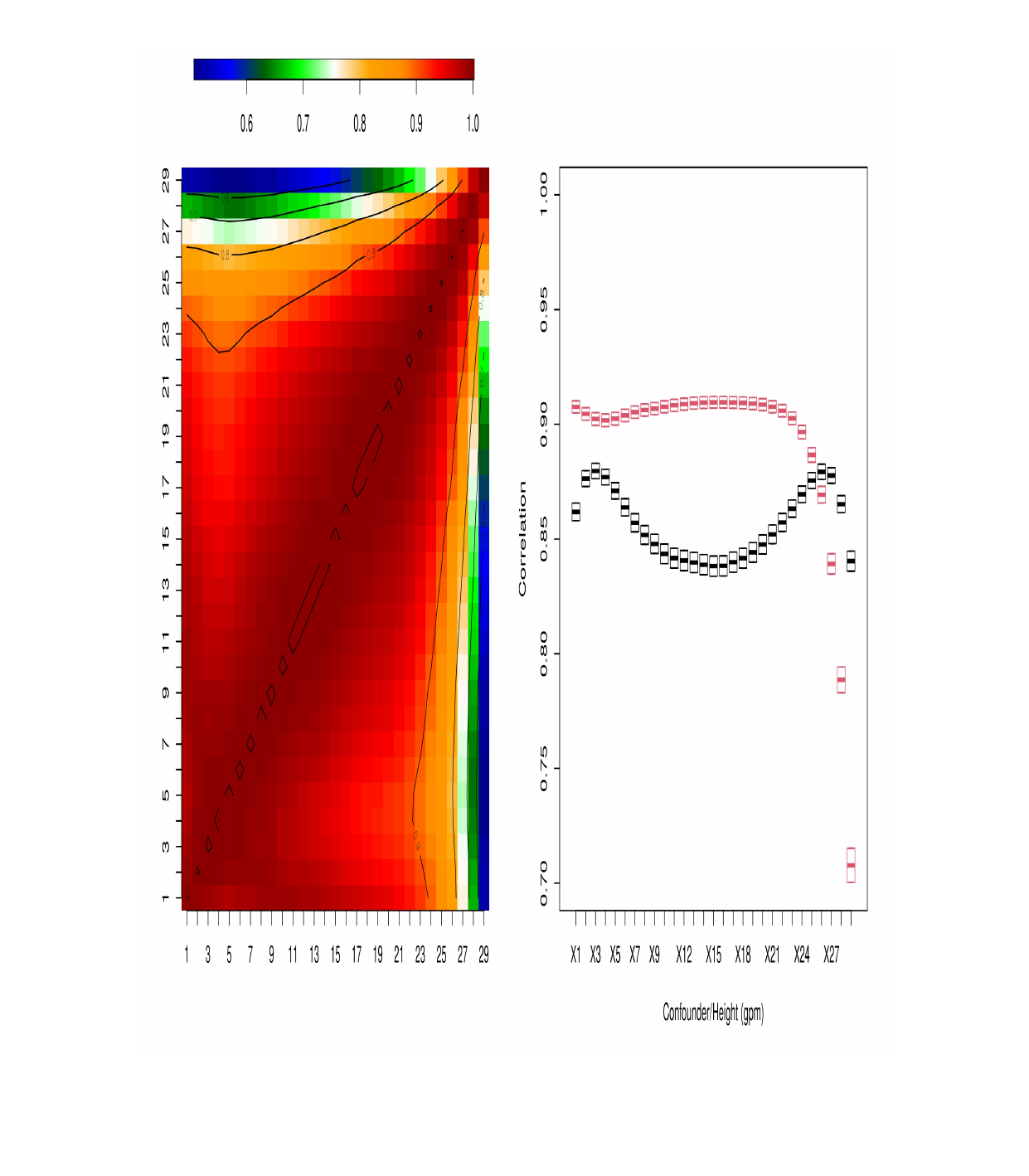}
\caption{Left: Correlation coefficient between all pairs of the 29 confounders (geopotential heights). 
The integers 1 through 29 displayed on the axes denote the pressure levels $100 hPa$ through 
$1000 hPa$, respectively. Right: Correlation between each of the 29 confounders (labeled on the x-axis) 
and the treatment variable (in black), and with the outcome variable (in red); shown are 95\% 
confidence intervals.}
\end{figure}

The choice of the two confounders to use for analysis can be narrowed down by another statistical
consideration; specifically, note that the graph in Figure 1 and the linear models in Equations (8)
and (9) presume a correlation between the confounder (geopotential height) and both the treatment 
variable (downward shortwave radiation flux) and the outcome (surface potential temperature).  
The right panel in Figure 2 shows the former (in black) and the latter (in red) for all 29
pressure levels labeled on the x-axis. As expected both correlations are relatively high across
most pressure levels, as reflected in the ``plateau'' in both curves. Two natural confounders are 
suggested by this result - one from the left side, and another from the right side of the plateau.
The correlation between outcome and confounder (red curve) falls off abruptly 
for pressure levels higher than $X_{24} (875 hPa)$, and so that is one of the confounders used here. 
The other pressure level, selected from the left side of the plateau, is $X_3 (150 hPa)$ for which the 
correlation between treatment and confounder (black curve) has a local maximum. Although the selection
criterion adopted here is not unique, this choice of the two pressure levels assures that two 
confounders are at significantly different pressure levels, and yet highly correlated with the 
treatment and outcome variables.

The histograms of the treatment (downward shortwave radiation flux), the outcome 
(surface potential temperature), and one of the two confounders (geopotential height at $150 hPa$) 
are shown in Figure 3. 
Also shown in Figure 3 are the pairwise scatterplots of these variables, confirming a linear association 
between them. Recall that the main goal of causal inference is to determine the extent of the 
correlation in the top scatterplot that is causal. 
The various peaks in these histograms and the non-random structure in the scatterplots
are a direct consequence of the spatial structure of the respective fields shown in Figure 4.

\begin{figure}[t]   % fig3
\center
\includegraphics[height=4in,width=5in]{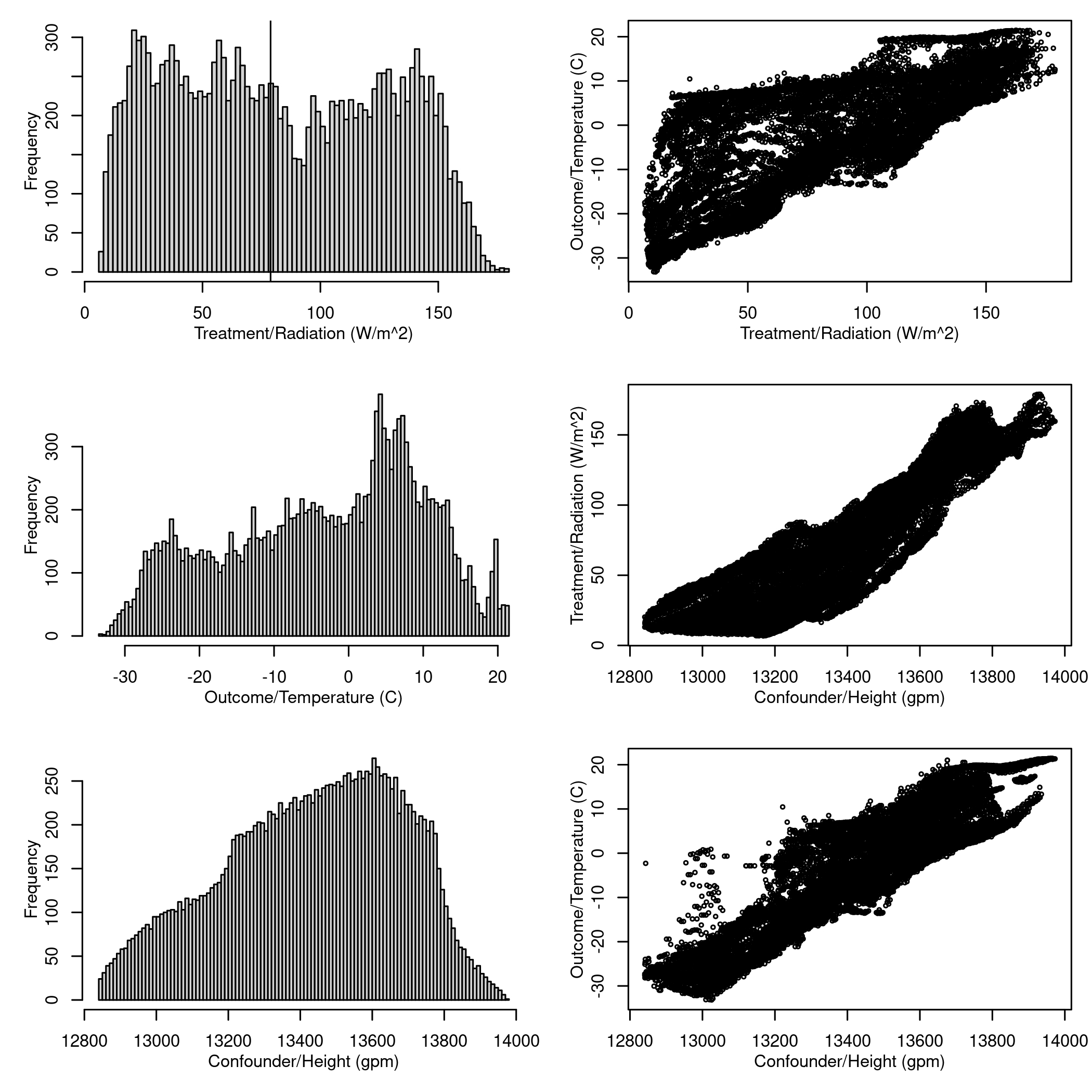}
\caption{Left: Histograms of the treatment variable (radiation), the outcome variable (temperature), 
and one confounder (height at $150 hPa$).  The vertical line on the first histogram denotes the median 
of radiation ($80 W/m^2$), used to dichotomize that variable. 
Right: pairwise scatterplots of the three variables, confirming linear associations between them.}

\end{figure}

\begin{figure}[t]   % fig4
\center
% \includegraphics[height=2in,width=2in]{map_radiation_dat4.jpg}
% \includegraphics[height=2in,width=2in]{map_temper_dat4.jpg} \\
% \center
% \includegraphics[height=2in,width=2in]{map_height_150_dat4.jpg}
% \includegraphics[height=2in,width=2in]{map_height_875_dat4.jpg}
 \includegraphics[height=6in,width=6in]{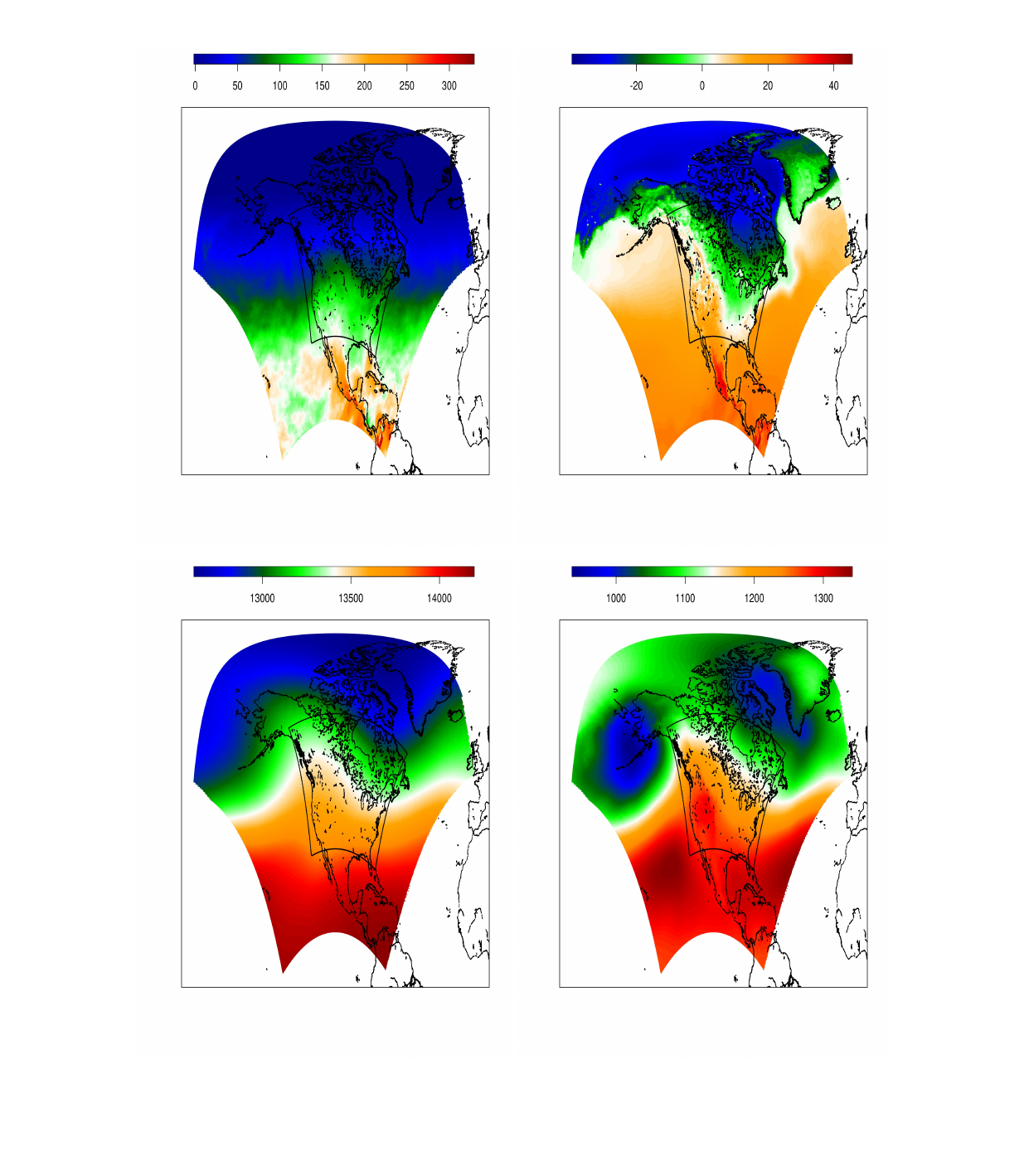}
\caption{Map of the treatment/radiation (top/left), outcome/temperature (top/right), and
the confounders/height at $150 hPa$ (bottom/left) and at $875 hPa$ (bottom/right). 
The inner domain shows the region of analysis. The median of radiation ($80 W/m^2$), used to 
dichotomize the treatment variable, appears at the boundary between green and blue colors.}
\end{figure}

Although all of the above considerations pertain to the treatment variable (downward shortwave 
radiation flux) as a continuous variable, the method described in this paper requires a binary 
treatment. To that end, at each grid point, 
the treatment variable is dichotomized by applying a threshold at the median of that variable. 
In other words, the treatment group consists of all grid points at which downward shortwave radiation 
flux exceeds its median across the entire spatial domain. The choice of the threshold is arbitrary and 
not based on a physical criterion, and so, other choices are possible. This median is shown as a 
vertical line in the histogram of radiation shown in Figure 3.

The bulk of the analysis is performed in R \citep{Rcore}, using the MatchIt package
\citep{greifer, ho_etal}. The main function
for developing the treatment model is called matchit(), and the function match.data() generates
the pseudo-population in which the confounders are balanced (i.e., have comparable histograms) across
the treatment and control groups. The outcome model is fitted by glm() and ATE is estimated
via the aforementioned g-computation method implemented in the function avg\_comparisons().

\section{Results}

Recall that the treatment group consists of all grid points at which downward shortwave radiation flux 
(radiation, for short) exceeds its median across the spatial domain. The outcome variable is surface 
potential temperature (temperature, for short), and the confounding variables are geopotential height 
(height, for short) at $150 hPa$ and $875 hPa$. To obtain a sense of the spatial variability
of ATE, it is estimated for each of 20 random samples of 8,000 grid points (about half) from the
$149 \times 117 = 17,433$ grid points in the spatial domain of analysis.  This sampling scheme also
has the added benefit of minimizing the effect of the spatial dependence of the data across the grid
points because spatial correlations generally fall off with distance.

To review, the matching method leads to a pseudo-population data set
in which the treatment and control groups are similar in terms of the confounders. As such,
the matching can be done for each of the confounders separately; but, as explained previously,
it is beneficial to perform the matching in terms of the PS. To that end, the treatment model
is developed, mapping the two confounders to the treatment. The predictions from this model
are estimates of PS. To assess balance, the SMD and Q-Q plots are computed both before and after 
matching.  If/when the PS, SMD, and Q-Q plots are deemed adequate, the ATE is estimated
for each of the 20 trials.

\subsection{Propensity Score (PS) Analysis}

The top panel in Figure 5 shows the histogram of the PS for the control group (in black) 
and the treatment group (in red). One desirable feature of these histograms is a right-tail for
one and a left-tail for the other, signifying that the predictors (confounders) in the treatment model,
Eq. (8), are good predictors of the response (downward shortwave radiation flux).\footnote{Given that 
a PS is a probability of
belonging to a treatment group, given the confounders, it may be useful to apply concepts
from the verification of probabilistic forecasts (e.g., attributes or reliability diagrams)
to better assess the quality of the treatment model.} Another desirable feature is the significant
overlap between the two conditional histograms; this important feature sets the stage for the
matching process, because it assures that PS can be used to match units in the control group
with other units in the treatment group.

\begin{figure}[t]  % fig5 made with matchit.R (not ps.R)
\center
\includegraphics[height=5in,width=4in]{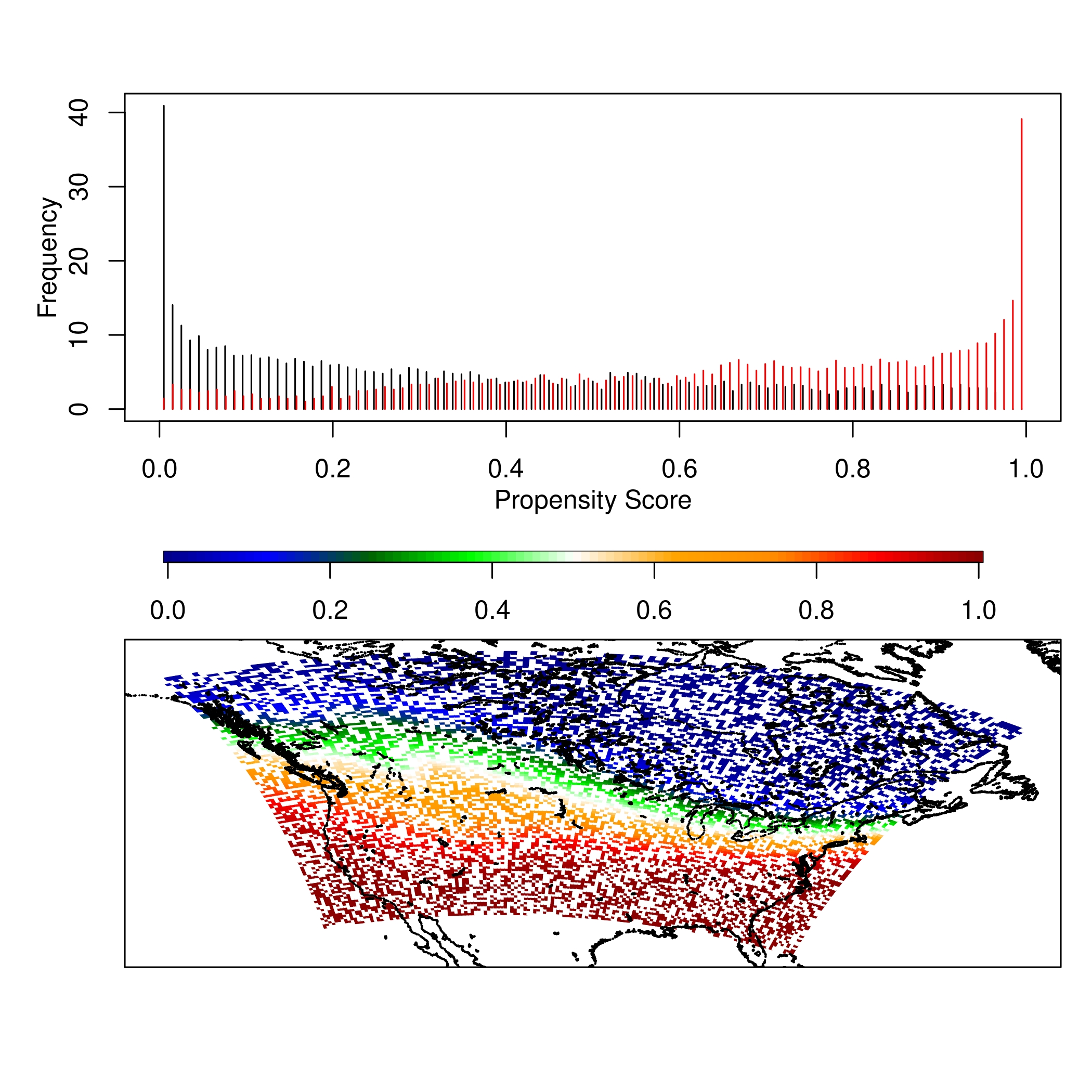}
\caption{Top: The histogram of PS for the control group (in black) and the treatment group (in red).
Bottom: The map of PS for the 8,000 grid points sampled for analysis.}
\end{figure}

The map of PS is shown in the bottom panel in Figure 5. The speckled nature of this map is a direct
consequence of the 8,000 points sampled from the domain. The pattern in this figure appears to reflect 
the latitude-dependent structure in the four fields shown in Figure 4. A more thorough comparison may 
provide a better understanding of the underlying relationships, and the role of topography in causal 
inference; see the discussion on spatial confounding in the discussion section. Also see the 
Closing Remarks for a discussion of yet another assumption, called {\it positivity.}

\subsection{Balance Analysis}

Figure 6 shows the boxplot of SMD values for PS and for each of the two confounders, based on the 
original data (i.e., before matching, in black) and for the pseudo-population (i.e., after matching by 
PS, in red). Each boxplot displays the variability due to the 20 aforementioned random samples. It can 
be seen that although matching by each confounder drastically improves the
SMD values, those values are still non-zero; this suggests that even in the matched data, there
are still some differences between the treatment and control groups in terms of each confounder.
However, the SMD values for PS are near-zero, implying that although each confounder individually 
does not balance the two groups, the PS estimated from both confounders leads to adequate balance.

\begin{figure}[t]   % fig6
\center
\includegraphics[height=3in,width=3in]{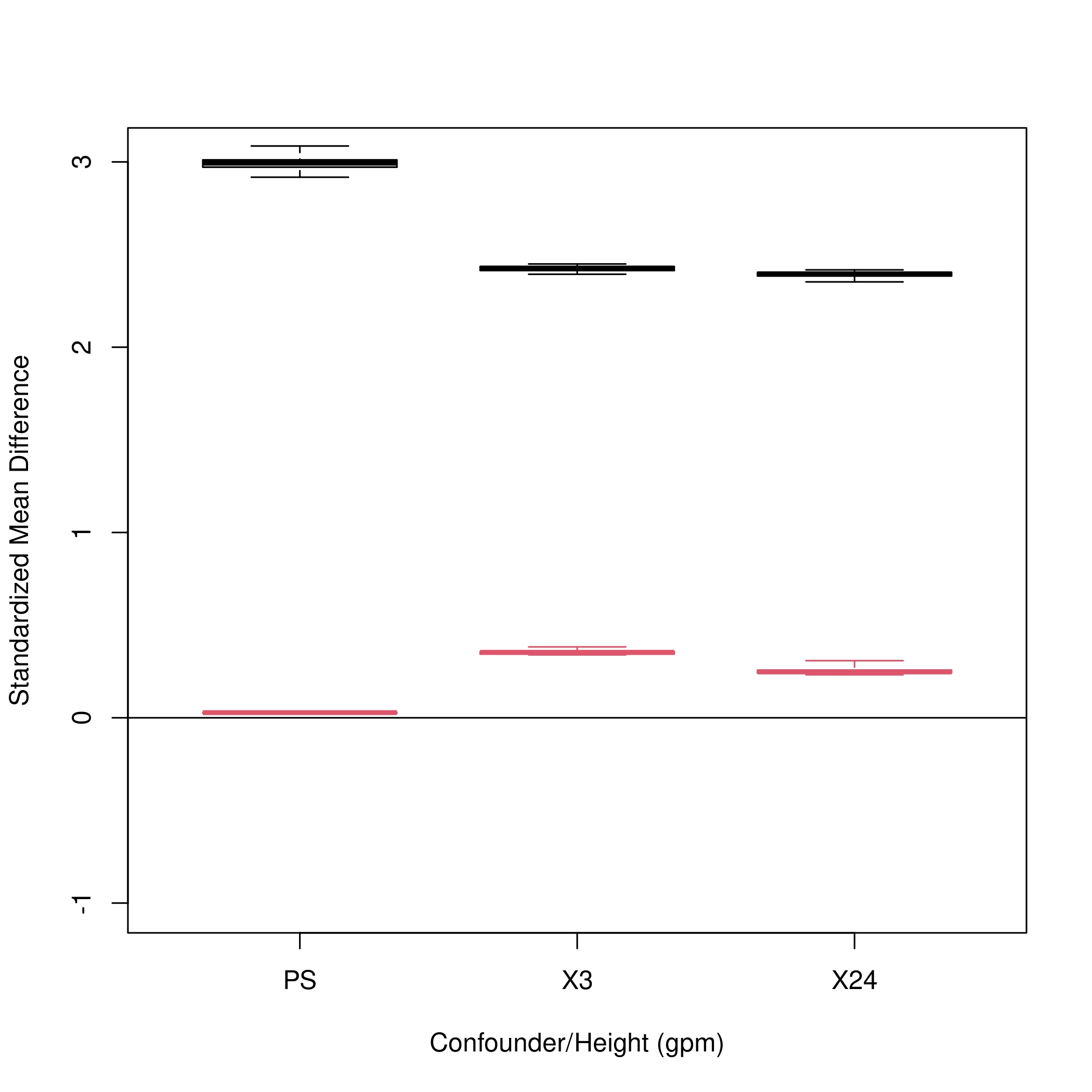}
\caption{The SMD for PS and each confounder, i.e., geopotential height at $150 hPa$ and $875 hPa$, 
respectively denoted $X_{3}$ and $X_{24}$. The black (red) boxplots are for
SMD before (after) matching. The boxplots display the variability of SMD across 20 random samples of
size 8,000 taken from the spatial domain of analysis.}
\end{figure}

As a mean, the SMD is a summary measure; as described in the Method section, a more diagnostic
assessment of balance is provided by two-sample Q-Q plots (Figure 7). The rows correspond to the
confounders. The left column shows the Q-Q plots before matching. The lack of overlap with
the diagonal (dashed) line suggests imbalance. For instance, as suggested by the strict, vertical
shift of the Q-Q plot with respect to the diagonal line, geopotential height in the treatment group
is generally larger than that in the control group. Said differently, before matching, for grid points
where downward shortwave radiation flux exceeds its median, the values of geopotential height
are generally larger than that over grid points where the flux is lower than its median.

\begin{figure}[t]  % fig7
\center
\includegraphics[height=5in,width=5in]{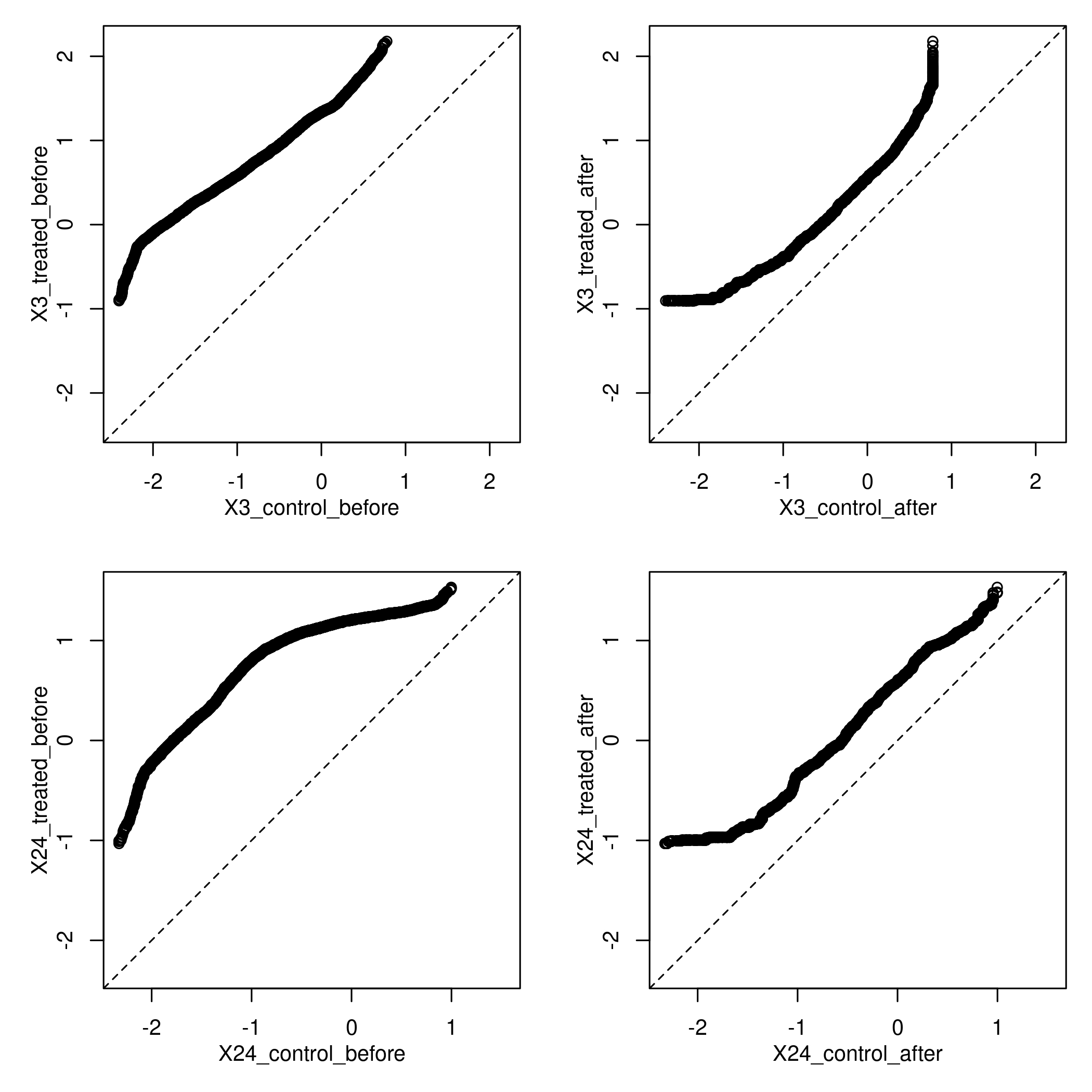}
\caption{Balance of the two covariates $X_{3}$ and $X_{24}$ (rows) assessed in terms of
two-sample Q-Q plots before matching (left column), and after matching (right column). The quantiles on
the x and y axes, correspond to the control and treatment groups, respectively.}
\end{figure}

This result is consistent with the basic meteorology of the spatial domain and time of year 
characterized in the selected data.
Greater values of the downward shortwave radiative fluxes tend to be present in the southern
portion of the domain due to the longer days and higher sun angles. In the middle and high latitude
portion of the domain there tends to be more frequent and stronger low-pressure disturbances
(storms) which are reflected by lesser values of mean lower-tropospheric geopotential heights and
associated with higher mean fractional cloud cover, in an overall sense. There are also longitudinal
differences in the shortwave radiative fluxes and geopotential heights in the domain, particularly
across the US. More specifically, in the monthly mean sense there are relatively high geopotential
heights over the western US resulting in suppressed cloudiness over the interior western US
extending eastward into the Great Plains states, with lower geopotential heights and enhanced
cloudiness over the eastern half of the US.\footnote{Cloud cover plays an important role in the
sister article.} In other words, the distribution of mean downward
shortwave radiative fluxes is related to both the shortwave flux at the top of the atmosphere and
mean weather patterns as represented by lower-tropospheric geopotential height patterns.

Furthermore, the nonlinear pattern in these Q-Q plots suggests that the histogram of the confounder
for the treatment group differs from that in the control group in ways other than a difference in
means and/or variances. Given these patterns, neither confounder is balanced across the two
treatment groups.  However, as seen in the right column in Figure 7, matching improves
balance in that the resulting Q-Q plots are more linear and closer to the diagonal line; see the 
Closing Remarks for a method of further impoving these results.
It is important to note that the Q-Q plot sugments where balance is improved for each confounder
(i.e., where the linear segments are closest to the diagonal line) are distinct.
In other words, each confounder balances the two groups differently, with the ultimate result
that the two confounders together lead to more balance via their contribution to PS, as
seen in the near-zero SMD values for PS in Figure 6.

\subsection{Estimation of Causal Effect}

The ultimate goal of causal inference is to estimate ATE. Figure 8 shows the 95\% confidence interval
(CI) for ATE using the matching method (in blue), for each of the 20 random samples, denoted as trials 
on the x-axis. As explained in the Method section, an alternative estimate of ATE can be obtained by 
adjusting for the effect of confounders in a traditional regression model that maps the treatment and 
the confounders to the outcome variable, without any matching weights. The CIs for these adjusted 
estimates are shown in red. Finally, for further comparison, also shown are the CIs for the simple 
difference in means (in black) based on the t-distribution.

Although there exists some variability across the trials, the conclusion is the same
across samples/trials: The simple difference in means yields relatively large, positive values for ATE. 
By contrast, the adjusted estimates are all negative. The ATE estimates based on matching 
are also negative, but smaller in magnitude than the adjusted estimate.
In short, there is evidence that downward shortwave radiation flux has a direct causal effect on 
surface potential temperature even after accounting for the confounding effect of geopotential height. 

\begin{figure}[t]  % fig8
\center
\includegraphics[height=3in,width=3in]{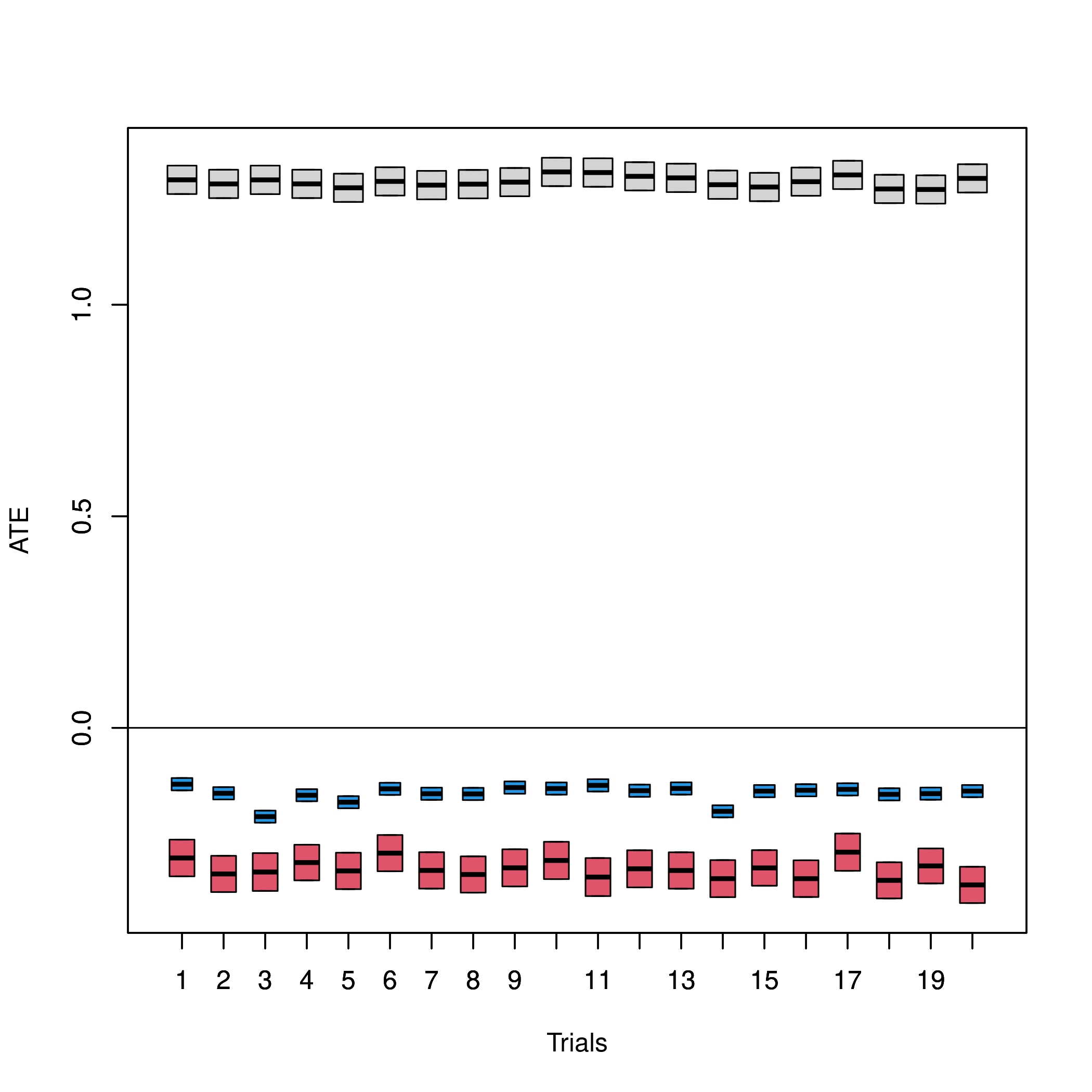}
\caption{The 95\% confidence interval for the simple difference in means (black),
for ATE in an adjusted model (red), and for ATE in a matched model (blue),
for 20 random samples of size 8,000 taken from the spatial domain of NARR.}
\end{figure}

It is worth pointing out that the effect is negative, i.e., increasing downward shortwave 
radiation flux causes a decrease in surface potential temperature, after the confounding effects
of geopotential height have been accounted for.  It is also worth noting that a simple comparison
of the mean of temperature across the control and treatment group suggests the ``wrong'' conclusion
about the direction of the relationship between downward shortwave radiation flux and surface 
temperature.

These conclusions must be tempered by all of the assumptions discussed in the Method section
and further addressed next.

% \conclusions  %% \conclusions[modified heading if necessary]
\conclusions[Conclusions and Discussion]

In assessing the direct, causal effect of one variable (treatment) on another (outcome), it is
important to account for the effect of other variables (confounders) on both the treatment
and the outcome.  Randomized controlled experiments reduce confounding by aiming to
assure that the treatment and the control groups are comparable in terms of all confounders.
The assessment of causality in observational studies, where a randomized experiment is not possible, has
given rise to a body of knowledge generally called causal inference. The abundance of observational
data in meteorology makes for a ripe environment for the application of causal inference methods.
This paper demonstrates one such method, based on regression (and matching), in which data on the
confounders are
fully available. An accompanying article demonstrates another regression-based method for situations
wherein data on the confounders are not available. Both methods are designed for a non-time-series
setting.

It is important to emphasize that the methods discussed in both papers are not designed to search
and/or identify causally related variables in data. Instead, the causal relationship between the
variables is established {\it a priori} based on theoretical considerations, and the causal inference
methods account for the effects of confounding variables that may influence that relationship.

Here, the method of matching is used to estimate the causal effect of downward shortwave radiation 
flux on surface potential temperature, when the confounding variables are geopotential height at 
different pressure levels.  For the sake of simplicity, only one month of NARR data is employed
in the analysis, and it is shown that the direct effect is significantly lower, and with the
opposite sign, compared to the simple difference in means. However, that estimate is smaller in
magnitude than the estimate that would follow from a traditional adjusted regression of surface potential 
temperature on downward shortwave radiation flux.

The interpretation of the ATE confidence intervals computed here deserves a comment.
The consistency assumption (Eq. 1) is more conveniently written as
$Y_i = A_i Y_i(1) + (1-A_i) Y_i(0)$, thereby making it clear that the outcome $Y_i$ is a
random variable because the treatment $A_i$ is a random variable.  This is an important point
because it implies that ideally (e.g., in a randomized controlled experiment) any variability in
estimates of ATE is due to the random assignment of the treatments, rather than from sampling
from a population \citep{athey_imbens}. Here, given the manner in which the
treatment and control grid points are defined, the confidence intervals reflect 
sampling variability across the spatial domain.

The reversal in the sign of ATE from positive to negative for the adjusted estimates can be
understood as an instance of Simpson's paradox \citep{pearl2019}, and is not a
consequence of causal inference. Although the paradox exists for both binary and continuous
variables, it is best demonstrated on the latter. As such, and only for this demonstration of
the paradox, consider the scatterplot of the outcome (surface potential temperature) versus the 
treatment (downward shortwave radiation flux) before the treatment is dichotomized, in the left panel
of Figure 9.  Clearly, there is a positive association between the two variables. To visualize the role
played by the confounder, the scatterplot is divided into 10 quantile intervals of the geopotential 
height at $150 hPa$, displayed as different colors in that scatterplot. Also shown are the regression 
fits to the data in each interval.  It can be seen that many of the intervals lead to negative slopes, 
regardless of the
positive slope of the overall scatterplot.  The right panel in Figure 9 shows the 95\% CI for the
slope of the regression fits in each interval. It can be seen that all of the values of geopotential 
height at $150 hPa$ between $13,250 gpm$ and $13,650 gpm$ have negative slopes.  Once again,
this sign reversal is a simple consequence of adjusting for confounders in a multiple regression
model for predicting the surface potential temperature from downward shortwave radiation
flux and geopotential height.

\begin{figure}[t]  % fig9
\center
\includegraphics[height=4in,width=7in]{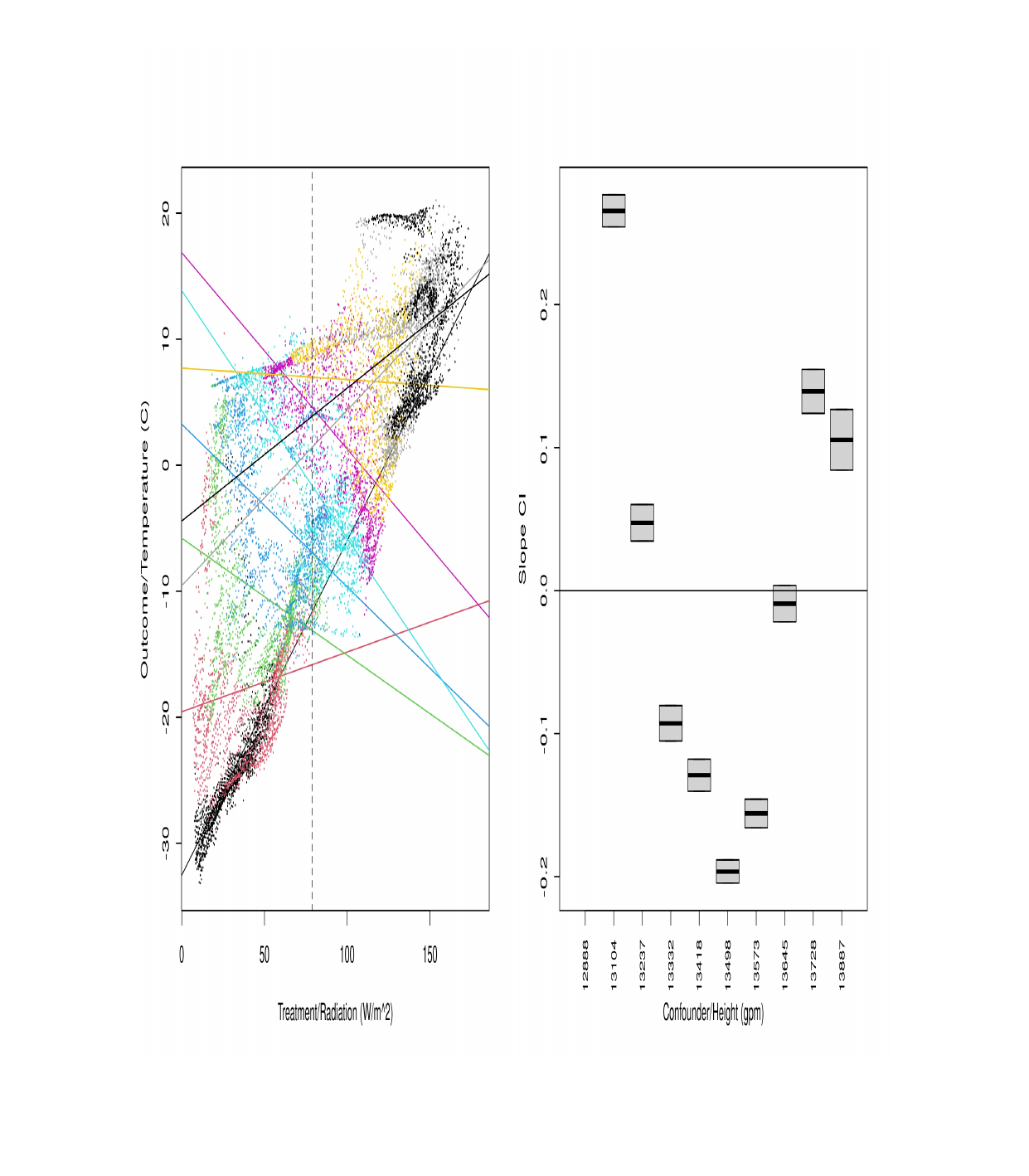}
\caption{Left: The scatterplot of outcome (temperature) versus treatment (radiation), before it
is dichotomized about a value of $80 W/m^2$ (the dashed, vertical line), with colors denoting 10 
quantile intervals of the confounder (geopotential height at $150 hPa$). Right: the 95\% confidence 
interval of the slope of the regression fits to each of the 10 intervals. The reversal of the sign 
of the slopes is an instance of Simpson's paradox.}
\end{figure}

The U-shaped pattern of slopes in the right panel of Figure 9 can be explained as follows:
The lower height and higher height portions of the $150 hPa$ range represent the southern/warmer 
and northern/colder portions of the analysis
domain where both the downward solar radiation and surface potential temperature systematically
decrease with latitude and hence are positively correlated. This relationship is
reinforced due to the correspondence between high values of downward shortwave radiation flux and 
surface potential temperature
over central Mexico where $150 hPa$ geopotential heights are relatively great. This contrasts with 
the middle latitude portion of the analysis domain where there are significant longitudinal
variations in both the downward solar radiation flux at the surface and in the surface
potential temperature, and where these variations do not align with one another.
More specifically, greater values of the radiation are present over the interior of the
western and the central US, i.e., essentially in the central longitudes of the domain,
while greater values of the temperature are centered to the west near the US west
coast. The result is the weakly negative correlations shown in Figure 9 for intermediate
values of the $150 hPa$ geopotential height.

The work here can be extended in a number of technical ways.
For example, here, the treatment model used for estimating PS is a generalized linear model (glm),
but the R function employed offers many alternatives (e.g., including generalized additive models,
gradient boosting machines, least absolute shrinkage and selection operator,
ridge regression, elastic net, classification trees, neural networks, random forests,
covariate balancing propensity score algorithm, and Bayesian additive regression trees).
Moreover, here, the matching is done with full matching of the propensity score, where all cases are
matched regardless of group membership. However, it is possible to match only treated units, match only
controls, use inverse probability weighting, or use nearest-neighbor matching \citep{lin_deng_fan}. Also, 
although consideration of nonlinear relations often falls in the purview of what is currently called 
Causal Machine Learning \citep{chernozhukov, feuerriegel}, many of the options available in the 
function matchit() accommodate nonlinear relationships.

Substantive extensions to the present work are also possible.  As explained above (in the theoretical
justification of the choice of variables), a more thorough analysis of the causal effect of downward
shortwave radiation flux on surface potential temperature requires the inclusion of other confounders
that account for seasonal variations, differences between land and water, zonal differences, topography, and more.

In order to focus on causal inference, the spatial structure of the fields has been ignored here.
One immediate consequence of the fact that the data at nearby grid points are correlated is that
the effective sample size is likely much smaller than the 8,000 grid points used here. Although
the resulting underestimation of the standard errors can have a significant effect on the conclusions
of a study, given the boxplots in Figure 6, it is unlikely that the conclusions would change
substantially if the boxplots displayed smaller variability resulting from a larger sample size. 
Regardless of the reduced effective sample size, it would be useful to assess the sensitivity of the 
results with respect to sample size similar to that in \citet{ombadi}. 

Ultimately, it will be important to account for spatial structure
more formally. One promising proposal capable of accounting for spatial structure is proposed by 
\citet{gilbert}. \citet{papadogeorgou} consider the case of binary treatment and outcome 
variables and have also developed a corresponding R package, ``geocausal." A related concept
arising from the existence of spatial structure is spatial confounding, where the relationship
between the treatment variable and the outcome variable is confounded by the fact that both are
affected by spatial location. Spatial confounding has been addressed by \citet{dupont1} and
\citet{dupont2}. A review of spatial causal inference can be found in \citep{reich}.

Temporal structures can also be accommodated. Although in Meteorology the phrase ``time series''
is used when referring to data with temporal structure, in Statistics the phrase panel-data is 
used for situations when time series are available for each unit. Extensions of the present 
methodology to panel-data can be found in the book by \citet{mixtape}, and the articles by 
\citet{xiao_wu} and \citet{athey}. It is worth mentioning that preprocessing steps aimed solely 
at removing temporal dependence (e.g., detrending or filtering) are not always appropriate, as 
they may alter the causal relationships of interest.

In summary, the method of causal inference described here allows one to assess the causal effect
of one variable on another, after the confounding effects of other variables have been taken
into account, all in observational data. The availability of large sets of gridded data in the 
atmospheric sciences allows for the application of this method to examine causal relations. 
The ubiquity of regression models in this method makes for a stepping stone for entering the field of 
causal inference.

\subsection{Closing Remarks}

Given the introductory nature of this (and the sister) article, several important details have been
addressed only superficially. However, those details are sufficiently important to warrant a
section dedicated to their discussion, hence these closing remarks.

\subsubsection{Assumptions}

All scientific models invoke assumptions. While some models may be robust with respect to violations
of the assumptions, others are not. In order to assess the sensitivity of the model to violations
of the assumptions, a body of literature has emerged, generally referred to as 
{\it Sensitivity Analysis} (SA); see \citep{cinelli_hazlett1, cinelli_hazlett2, oganisian} for an example of 
SA in causal inference. 

Although this article does not delve into SA in order to focus on the basics of causal inference, some 
violations of assumptions are evident. This section delves deeper into three of them, which are
likely violated to some extent.

For example, consider NUCA, mentioned in the Method section; 
effectively, it assumes that there are no unmeasured confounders, or equivalently that all confounders 
are included as covariates in the treatment model (Eq. (8)). Given the complex processes underlying
the relationship between downward shortwave radiation flux and surface potential temperature, it is 
likely that this assumption is violated. As such, the estimate of ATE obtained here may be biased.

Returning to Figure 5, technically, the existence of the singular peaks at PS of 0 and 1 is 
undesirable, because it implies
that some units will be impossible to match in the matching process. Indeed, a PS of 0 or 1 is said
to violate another assumption of causal inference, called {\it positivity}; for the consequences of
this violation, which extend beyond the matching method, see \citep{kanga_positivity}. A common
practice in such situations is to exclude the units with PS of 0 or 1, in which case the estimated
ATE applies only to the included units. Doing so leads to the exclusion of the deep-red and deep-blue 
grid points in the map of PS in Figure 5; it also leads to Q-Q plots (e.g., in Figure 7) that
have sigificantly more overlap with the diagonal line. For the
present demonstration, these points are not excluded, not
only to simplify the presentation, but also because excluding the points does not significantly
affect the estimates of ATE.\footnote{The R function for matching used here, has an argument that allows
for the exclusion of units that violate this assumption; discard = ``none'' does not exclude any units,
while discard = ``both'' excludes units from the treatment and control groups.}

Finally, there exists another pair of assumptions that deserve mention; 1) the potential outcome for one
unit is not affected by the treatment at a different unit, and 2) there are no multiple versions of
the treatment. This pair of assumptions are grouped under the single name Stable Unit Treatment 
Assumption (SUTVA). The former is generally called the no-interference assumption (or no spillover 
effect).\footnote{The name associated with the latter assumption is inconsistent across authors; 
some (e.g., \citet{ding_book}) refer to it as ``consistency,'' an unfortunate choice since other
authors (e.g., \citet{hernan_whatif}) use that name in reference to the assumption in Eq. (1).
The reason for the naming inconsistency is that there is a connection between the assumption in Eq. 1 
and the assumption that there are no multiple versions of the treatment. For example, 
\citet{vanderweele_consistency} shows that multiple versions of treatment may be allowed
as long as they all result in the same outcome; but in doing so, the statement in Eq. (1) requires
revision.} SUTVA is an important assumption not only in causal
inference from observational data, but even in randomized experiments. 

The first (no-interference) assumption in SUTVA is implicit in the very notation for potential 
outcomes - $Y_i(1), Y_i(0)$ - which is also why \citep{vanderweele_consistency} found it necessary to 
generalize that notation in order to accommodate violations of the second part of the SUTVA assumption.
In the present application, the no-interference assumption translates to the requirement that the 
surface potential temperature at the $i^{th}$ grid point does not depend on downward shortwave 
radiation flux at another grid point. This assumption is probably violated not only across
nearby units (due to spatial correlations), but also across distant units due to complex
atmospheric patterns. Said differently, interference is not synonymous with spatial correlation, 
because interference can lead to spatial correlations, but not all spatial correlations are a result of 
interference. It could be argued that in the present application the spatial correlations are not a
consequence of interference, but rather the effect of a latent variable, namely location. One
may also argue that the spatial correlations are indeed a consequence of interference. Given
such ambiguities, it is worthwhile to examine methods that account for spatial confounding,
e.g., \citep{gilbert} and \citep{reich} and methods that aim to deal with interference directly, 
e.g., \citep{hudgens} and \citep{savje}.

\subsubsection{Intervention}

As mentioned at the outset, there is some controversy in the field regarding the mantra
``No Causation without Intervention." This article takes no sides on that issue. Suffice it to say
that there is variability across fields in the interpretation of ``intervention." For example,
\citet{dupont2} consider the causal effect of monthly mean precipitation on monthly mean temperature 
at 2m. As another example, \citet{galaxy} examine the effect of internal and external process in the 
formation of galaxies.  It is difficult to imagine a thought-experiment wherein one can intervene on 
the respective treatments in these studies in any realistic (physical) manner. In the present
application, however, the existence of sophisticated numerical models does allow for some level
of intervention.

\codeavailability{R Code for the analysis performed here is available as supplementary material} %% use this section when having only software code available

\dataavailability{The NARR data are publicly available at the following URL:\\
https://www.ncei.noaa.gov/products/weather-climate-models/north-american-regional.} %% use this section when having only data sets available

% \codedataavailability{TEXT} %% use this section when having data sets and software code available

% \sampleavailability{TEXT} %% use this section when having geoscientific samples available

% \videosupplement{TEXT} %% use this section when having video supplements available

\appendix
\section{Glossary of Causal Inference}    %% Appendix A

This appendix provides a partial list of terms and phrases commonly used in causal inference.

Treatment Variable: The variable whose causal effect on the outcome variable is being assessed.

Treatment and control: The two levels of a binary treatment variable.

Outcome Variable: The variable of interest being affected by the treatment variable.

Lurking variable: A generic name for a variable that is associated with both the treatment variable and the outcome variable.

Confounding Variable: A specific type of a lurking variable wherein it causally affects both the treatment variable and the outcome variable.

Potential Outcome: The outcome variable for a unit under a potential treatment value.

Counterfactual Outcome: The hypothetical outcome that would have occurred if a unit had
received a different treatment than it actually did.

Unit: The entity to which the treatment is applied.

Average Treatment Effect (ATE): The difference between the outcome variable under the two treatment levels, averaged across all units.

ATC: The average treatment effect in the control group.

ATT: The average treatment effect in the treatment group.

Standardized Mean Difference (SMD):  The difference of two conditional means, divided by the pooled standard deviation.

Propensity Score: The probability of being selected into the treatment group, given some set of covariates.

Balance: When treatment and control groups are similar in terms of confounders.

Matching: The process of pairing units in the treatment group with those in the control group
based on similarities in confounders or in propensity score.

The Treatment Model: A regression model for estimating the propensity score.

The Outcome Model: A regression model for estimating the potential outcome.

\section{Derivations and Proofs}

% \subsection{}     %% Appendix A1, A2, etc.

This appendix provides a derivation of Eq. 3 and a proof of two assertions made in the Method section.

Recall the definition $ATE=E[Y(1) - Y(0)]$, and $\pi$ as the probability of a unit
receiving the treatment, and note that
\begin{align*}
ATE &= \pi\cdot \left(E[Y(1)|A=1] - E[Y(0)|A=1] \right) + (1-\pi)\cdot \left(E[Y(1)|A=0] - E[Y(0)|A=0] \right) \\
&= E[Y(1)|A=1] - E[Y(0)|A=0] - (1-\pi)\cdot E[Y(1)|A=1] - \pi \cdot E[Y(0)|A=1] \\
&\quad + (1-\pi)\cdot E[Y(1)|A=0] + \pi\cdot E[Y(0)|A=0]\\
&= E[Y(1)|A=1] - E[Y(0)|A=0] - (1-\pi)\cdot \left(E[Y(1)|A=1] - E[Y(0)|A=1] \right) \\
&\quad + (1-\pi)\left(E[Y(1)|A=0] - E[Y(0)|A=0] \right) - \left(E[Y(0)|A=1] - E[Y(0)|A=0] \right)\\
&= E[Y(1)|A=1] - E[Y(0)|A=0] \\
&\quad - (1-\pi)\cdot \left\{\underbrace{\left(E[Y(1)|A=1] - E[Y(0)|A=1]\right)}_{ATT} - \underbrace{\left(E[Y(1)|A=0] - E[Y(0)|A=0]\right)}_{ATC} \right\}\\
&\quad - \left(E[Y(0)|A=1] - E[Y(0)|A=0] \right)\\
&= E[Y(1)|A=1] - E[Y(0)|A=0] - (1-\pi)(ATT - ATC) - \left(E[Y(0)|A=1] - E[Y(0)|A=0] \right).
\end{align*}
Eq. (3) follows by rearrangement of terms.

The remainder of this Appendix provides a proof of the following assertions:
\begin{eqnarray}
A \perp X \;|\; p(X), \\
A \perp (Y(0), Y(1)) \;|\; X \implies A \perp (Y(0), Y(1)) \;|\; p(X),
\end{eqnarray}
where $p(X) = pr(A=1 | X)$ is called the propensity score.
In words, the propensity score is a balancing score, and if conditioning on $X$ renders the treatment
independent of the potential outcomes, then so does conditioning on the propensity score
(Cunningham 2021, Deng 2021).

Note that the propensity score can be written as $E[A|X]$, because $A$ is binary. Then,
\begin{equation}
pr(A=1 \;|\; X, p(X)) = E[A \;|\; X, p(X)] = E[A \;|\; X] = p(X),
\end{equation}
where the second equality follows because conditioning on $X$ renders conditioning on $X$ and $p(X)$
redundant. Similarly,
\begin{eqnarray}
pr(A=1 \;|\; p(X)) &=& E[A \;|\; p(X)] \\
                  &=& E \bigl[ E[A \;|\; X, p(X)] \;|\; p(X) \bigr] \\
                  &=& E \bigl[ E[A \;|\; X] \;|\; p(X) \bigr]\\
                  &=& E[ p(X) \;|\; p(X)] = p(X).
\end{eqnarray}
Equation (B5) follows from (B4) because of the law of iterated expectation in probability theory.
Equation (B6) follows from (B5) because conditioning on $X$ renders conditioning on $X$ and $p(X)$
redundant, and the last equality follows from a property of the conditional expectation in
probability theory.

Equations (B3) and (B7) imply $pr(A=1|X,p(X))=pr(A=1|p(X))$, and hence, the assertion (B1). Similarly,
\begin{eqnarray}
pr(A=1 \;|\; Y(0), Y(1), p(X)) &=& E[A \;|\; Y(0), Y(1), p(X)] \\
                           &=& E \bigl[ E[A \;|\; Y(0), Y(1), X, p(X)] \;|\; Y(0), Y(1), p(X) \bigr] \\
                           &=& E \bigl[ E[A \;|\; Y(0), Y(1), X] \;|\; Y(0), Y(1), p(X) \bigr]\\
                           &=& E \bigl[ E[A \;|\; X] \;|\; Y(0), Y(1), p(X) \bigr]\\
                           &=& E[ p(X) \;|\; Y(0), Y(1), p(X)] = p(X).
\end{eqnarray}
Eq. (B9) follows from (B8) from the iterated law of expectations. The conditional independence of the
treatment and the potential outcomes, given the covariate, is used in going from (B10) to (B11).
From (B7) and (B12) it follows that $pr(A=1 | Y(0), Y(1), p(X)) = pr(A=1 | p(X))$, and hence, the
assertion (B2).

\noappendix       %% use this to mark the end of the appendix section.

%%%%%%%%%%%%%%%%%%%%%%%%%%%%%%%%%%%%%%%%%%%%%%%%%%%%%%%%%%%%%%%%%%%%%%%%%%%%%%%

%% Regarding figures and tables in appendices, the following two options are possible depending on your general handling of figures and tables in the manuscript environment:

%% Option 1: If you sorted all figures and tables into the sections of the text, please also sort the appendix figures and appendix tables into the respective appendix sections.
%% They will be correctly named automatically.

%% Option 2: If you put all figures after the reference list, please insert appendix tables and figures after the normal tables and figures.
%% To rename them correctly to A1, A2, etc., please add the following commands in front of them:

\appendixfigures  %% needs to be added in front of appendix figures

\appendixtables   %% needs to be added in front of appendix tables

%% Please add \clearpage between each table and/or figure. Further guidelines on figures and tables can be found below.

%%%%%%%%%%%%%%%%%%%%%%%%%%%%%%%%%%%%%%%%%%%%%%%%%%%%%%%%%%%%%%%%%%%%%%%%%%%%%%%%%%%

\authorcontribution{Marzban and Zhang have contributed to the causal inference aspects of the work, and Bond and Richman have provided meteorological expertise.} %% mandatory

\competinginterests{The authors declare that they have no conflict of interest.} %% mandatory 

\disclaimer{No funding has been provided for this work.} %% optional section

\begin{acknowledgements}
None.
\end{acknowledgements}

%% REFERENCES

%% The reference list is compiled as follows:

% \begin{thebibliography}{}
% \bibitem[AUTHOR(YEAR)]{LABEL1}
% REFERENCE 1
% \bibitem[AUTHOR(YEAR)]{LABEL2}
% REFERENCE 2
% \end{thebibliography}

%% Since the Copernicus LaTeX package includes the BibTeX style file copernicus.bst,
%% authors experienced with BibTeX only have to include the following two lines:
%%
 \bibliographystyle{copernicus}
 \bibliography{paper.bib}

\begin{thebibliography}{82}
\providecommand{\natexlab}[1]{#1}
\providecommand{\url}[1]{\texttt{#1}}
\providecommand{\urlprefix}{}
\expandafter\ifx\csname urlstyle\endcsname\relax
  \providecommand{\doi}[1]{https://doi.org/\discretionary{}{}{}#1}\else
  \providecommand{\doi}{https://doi.org/\discretionary{}{}{}\begingroup
  \urlstyle{rm}\Url}\fi

\bibitem[{Abadie and Imbens(2016)}]{abadie_imbens}
Abadie, A. and Imbens, G.~W.: Matching on the estimated propensity score,
  Econometrica, 63, 781--807, 2016.

\bibitem[{Athey and Imbens(2017)}]{athey_imbens}
Athey, S. and Imbens, G.~W.: The econometrics of randomized experiments, in:
  Handbook of Economic Field Experiments, vol.~1, pp. 73--140, Elsevier, 2017.

\bibitem[{Athey et~al.(2021)Athey, Bayati, Doudchenko, Imbens, and
  Khosravi}]{athey}
Athey, S., Bayati, M., Doudchenko, N., Imbens, G., and Khosravi, K.: Matrix
  Completion Methods for Causal Panel Data Models, Journal of the American
  Statistical Association, 116, 1716--1730,
  \doi{10.1080/01621459.2021.1891924}, 2021.

\bibitem[{Camps-Valls et~al.(2023)Camps-Valls, Gerhardus, Urman, Varando,
  Martius, Balaguer-Ballester, Vinuesa, Diaz, Zanna, and Runge}]{camps}
Camps-Valls, G., Gerhardus, A., Urman, N., Varando, G., Martius, G.,
  Balaguer-Ballester, E., Vinuesa, R., Diaz, E., Zanna, L., and Runge, J.:
  Discovering causal relations and equations from data, Physics Reports, 1044,
  1--68, 2023.

\bibitem[{Chattopadhyay and Zubizarreta(2023)}]{chatto}
Chattopadhyay, A. and Zubizarreta, J.: On the implied weights of linear
  regression for causal inference, Biometrika, 110, 615--629, 2023.

\bibitem[{Chernozhukov et~al.(2018)Chernozhukov, Chetverikov, Demirer, Duflo,
  Hansen, Newey, and Robins}]{chernozhukov}
Chernozhukov, V., Chetverikov, D., Demirer, M., Duflo, E., Hansen, C., Newey,
  W., and Robins, J.: Double/debiased machine learning for treatment and
  structural parameters, The Econometrics Journal, 21, C1--C68,
  \doi{10.1111/ectj.12097}, 2018.

\bibitem[{Cinelli and Hazlett(2019)}]{cinelli_hazlett1}
Cinelli, C. and Hazlett, C.: Making sense of sensitivity: extending omitted
  variable bias, Journal of the Royal Statistical Society: Series B
  (Statistical Methodology), 82,
  \urlprefix\url{https://api.semanticscholar.org/CorpusID:189811135}, 2019.

\bibitem[{Cinelli and Hazlett(2025)}]{cinelli_hazlett2}
Cinelli, C. and Hazlett, C.: An omitted variable bias framework for sensitivity
  analysis of instrumental variables, Biometrika, 112, asaf004,
  \doi{10.1093/biomet/asaf004}, 2025.

\bibitem[{Cochran(1965)}]{cochran}
Cochran, W.~G.: The Planning of Observational Studies of Human Populations,
  Journal of the Royal Statistical Society, Series A (General), 128, 234--266,
  \doi{https://doi.org/10.2307/2344179}, 1965.

\bibitem[{Cole and Frangakis(2009)}]{cole_frangakis}
Cole, S.~R. and Frangakis, C.~E.: The consistency statement in causal
  inference: a definition or an assumption?, Epidemiology, 20, 3--5,
  \doi{10.1097/EDE.0b013e31818ef366}, 2009.

\bibitem[{Cunningham(2021)}]{mixtape}
Cunningham, S.: Causal Inference: The Mixtape, Yale University Press, New Haven
  and London, ISBN 9780300251685, 2021.

\bibitem[{Dawid(1979)}]{dawid}
Dawid, A.~P.: Conditional independence in statistical theory, Journal of the
  Royal Statistical Society: Series B (Statistical Methodology), 41, 1--15,
  1979.

\bibitem[{Dehejia and Wahba(1999)}]{dehejia}
Dehejia, R.~H. and Wahba, S.: Causal Effects in Nonexperimental Studies:
  Reevaluating the Evaluation of Training Programs, Journal of the American
  Statistical Association, 94, 1053--1062,
  \urlprefix\url{http://www.jstor.org/stable/2669919}, 1999.

\bibitem[{Ding(2024)}]{ding_book}
Ding, P.: A First Course in Causal Inference, Chapman \& Hall/CRC Texts in
  Statistical Science, 2024.

\bibitem[{Dorn(1953)}]{dorn}
Dorn, H.~F.: Philosophy of inferences from retrospective studies, Am J Public
  Health and the Nation's Health, 43, 677--683,
  \doi{10.2105/ajph.43.6_pt_1.677}, 1953.

\bibitem[{Dupont et~al.(2022)Dupont, Wood, and Augustin}]{dupont1}
Dupont, E., Wood, S.~N., and Augustin, N.~H.: Spatial+: A novel approach to
  spatial confounding, Biometrics, 78, 1279–1290, \doi{10.1111/biom.13656},
  2022.

\bibitem[{Dupont et~al.(2025)Dupont, Marques, and Kneib}]{dupont2}
Dupont, E., Marques, I., and Kneib, T.: Demystifying Spatial Confounding,
  \urlprefix\url{https://arxiv.org/abs/2309.16861}, 2025.

\bibitem[{Ebert-Uphoff and Deng(2012)}]{ebert_deng}
Ebert-Uphoff, I. and Deng, Y.: Causal discovery for climate research using
  graphical models, Journal of Climate, 25, 5648--5665,
  \doi{10.1175/JCLI-D-11-00387.1}, 2012.

\bibitem[{Feuerriegel et~al.(2024)Feuerriegel, Frauen, Melnychuk, Schweisthal,
  Hess, Curth, Bauer, Kilbertus, Kohane, and van~der Schaar}]{feuerriegel}
Feuerriegel, S., Frauen, D., Melnychuk, V., Schweisthal, J., Hess, K., Curth,
  A., Bauer, S., Kilbertus, N., Kohane, I.~S., and van~der Schaar, M.: Causal
  machine learning for predicting treatment outcomes, Nature Medicine, 30,
  958--968, 2024.

\bibitem[{Fisher(1935)}]{fisher}
Fisher, R.~A.: The Design of Experiments, Oliver and Boyd, Edinburgh and
  London, 1st edn., 1935.

\bibitem[{Gelman(2013)}]{gelman_blog}
Gelman, A.: Post on the controversial claim that high or low genetic diversity
  is bad for the economy, \url{
  https://statmodeling.stat.columbia.edu/2013/01/10/that-controversial-claim-that-high-genetic-diversity-or-low-genetic-diversity-is-bad-for-the-economy/
  }, 2013.

\bibitem[{Gelman and Hill(2007)}]{gelman_book}
Gelman, A. and Hill, J.: Data Analysis Using Regression and
  Multilevel/Hierarchical Models, Cambridge University Press, 2007.

\bibitem[{Gilbert et~al.(2021)Gilbert, Datta, Casey, and Ogburn}]{gilbert}
Gilbert, B., Datta, A., Casey, J.~A., and Ogburn, E.~L.: A causal inference
  framework for spatial confounding, arXiv preprint arXiv:2112.14946, 2021.

\bibitem[{Granger(2004)}]{granger}
Granger, C. W.~J.: Time series analysis, cointegration, and applications,
  American Economic Review, 94, 421--425, \doi{10.1257/0002828041464669}, 2004.

\bibitem[{Greifer(2023)}]{greifer}
Greifer, N.: Estimating effects using {MatchIt},
  \url{https://cran.r-project.org/web/packages/MatchIt/vignettes/estimating-effects.html},
  2023.

\bibitem[{Greifer and Stuart(2021)}]{greifer_stuart}
Greifer, N. and Stuart, E.~A.: Matching methods for confounder adjustment: An
  addition to the epidemiologist’s toolbox, Epidemiologic Reviews, 43,
  118--129, \doi{10.1093/epirev/mxab003}, 2021.

\bibitem[{Hannart(2019)}]{hannart2019}
Hannart, A.: An improved projection of climate observations for detection and
  attribution, Advances in Statistical Climatology, Meteorology and
  Oceanography, 5, 161--171, \doi{10.5194/ascmo-5-161-2019}, 2019.

\bibitem[{Hannart et~al.(2016)Hannart, Pearl, Otto, Naveau, and
  Ghil}]{hannart2016}
Hannart, A., Pearl, J., Otto, F. E.~L., Naveau, P., and Ghil, M.: Causal
  counterfactual theory for the attribution of weather and climate-related
  events, Bulletin of the American Meteorological Society, 97, 99--110,
  \doi{10.1175/BAMS-D-14-00034.1}, 2016.

\bibitem[{Hansen and Hurwitz(1943)}]{hansen-hurwitz}
Hansen, M.~H. and Hurwitz, W.~N.: On the theory of sampling from finite
  populations, The Annals of Mathematical Statistics, 14 (4), 333--362, 1943.

\bibitem[{Heinrich et~al.(2010)Heinrich, Maffioli, and Vazquez}]{heinrich}
Heinrich, C., Maffioli, A., and Vazquez, G.: A primer for applying
  propensity-score matching, Technical Note IDB-TN-161, Inter-American
  Development Bank, Office of Strategic Planning and Development Effectiveness,
  2010.

\bibitem[{Hern{\'a}n and Robins(2020)}]{hernan_whatif}
Hern{\'a}n, M.~A. and Robins, J.~M.: Causal Inference: What If, Chapman \&
  Hall/CRC, Boca Raton, 2020.

\bibitem[{Hirano and Imbens(2004)}]{hirano_imbens}
Hirano, K. and Imbens, G.~W.: The Propensity Score with Continuous Treatments,
  chap.~7, pp. 73--84, John Wiley \& Sons, Ltd, ISBN 9780470090459,
  \doi{https://doi.org/10.1002/0470090456.ch7}, 2004.

\bibitem[{Hirt et~al.(2020)Hirt, Craig, Sch{\"a}fer, Savre, and Heinze}]{hirt}
Hirt, M., Craig, G.~C., Sch{\"a}fer, S. A.~K., Savre, J., and Heinze, R.:
  Cold-pool-driven convective initiation: Using causal graph analysis to
  determine what convection-permitting models are missing, Quarterly Journal of
  the Royal Meteorological Society, 146, 2205--2227, \doi{10.1002/qj.3788},
  2020.

\bibitem[{Ho et~al.(2007)Ho, Imai, King, and Stuart}]{ho_etal}
Ho, D.~E., Imai, K., King, G., and Stuart, E.~A.: Matching as nonparametric
  preprocessing for reducing model dependence in parametric causal inference,
  Political Analysis, 15, 199--236, \doi{10.1093/pan/mpl013}, 2007.

\bibitem[{Holland(1986)}]{holland}
Holland, P.~W.: Statistics and causal inference, Journal of the American
  Statistical Association, 81, 945--960, 1986.

\bibitem[{Hudgens and Halloran(2008)}]{hudgens}
Hudgens, M.~G. and Halloran, M.~E.: Toward causal inference with interference,
  Journal of the American Statistical Association, 103, 832--842, 2008.

\bibitem[{Imai and Dyk(2004)}]{kosuke_vandyke}
Imai, K. and Dyk, D. A.~V.: Causal inference with general treatment regimes:
  Generalizing the propensity score, Journal of the American Statistical
  Association, 99.467, 854--866, 2004.

\bibitem[{Imbens and Rubin(2015)}]{imbens_rubin}
Imbens, G.~W. and Rubin, D.~B.: Causal Inference in Statistics, Social, and
  Biomedical Sciences, Cambridge University Press, 2015.

\bibitem[{Imbens and Xu(2024)}]{imbens_xu}
Imbens, G.~W. and Xu, T.: Lessons Learned,
  \url{http://dx.doi.org/10.2139/ssrn.4849285}, 2024.

\bibitem[{Kanga et~al.(2016)Kanga, Chan, and Mi-Ok~Kim}]{kanga_positivity}
Kanga, J., Chan, W., and Mi-Ok~Kim, P. M.~S.: Practice of causal inference with
  the propensity of being zero or one: assessing the effect of arbitrary
  cutoffs of propensity scores, Communications for Statistical Applications and
  Methods, 23, 1--20, 2016.

\bibitem[{Kretschmer et~al.(2021)Kretschmer, Adams, Arribas, Prudden, Robinson,
  Saggioro, and Shepherd}]{kretschmer}
Kretschmer, M., Adams, S.~V., Arribas, A., Prudden, R., Robinson, N., Saggioro,
  E., and Shepherd, T.~G.: Quantifying causal pathways of teleconnections,
  Bulletin of the American Meteorological Society, pp. E2247--E2263,
  \doi{10.1175/BAMS-D-20-0117.1}, 2021.

\bibitem[{LaLonde(1986)}]{lalonde}
LaLonde, R.~J.: Evaluating the Econometric Evaluations of Training Programs
  with Experimental Data, The American Economic Review, 76, 604--620,
  \urlprefix\url{http://www.jstor.org/stable/1806062}, 1986.

\bibitem[{Li and Chu(2023)}]{li_chu}
Li, S. and Chu, Z., eds.: Machine Learning for Causal Inference, Springer,
  \doi{10.1007/978-3-031-35051-1_14}, 2023.

\bibitem[{Lin et~al.(2023)Lin, Ding, and Han}]{lin_deng_fan}
Lin, Z., Ding, P., and Han, F.: Estimation Based on Nearest Neighbor Matching:
  From Density Ratio to Average Treatment Effect, Econometrica, 91, 2187--2217,
  \doi{10.3982/ECTA20598}, 2023.

\bibitem[{Marzban et~al.(2026)Marzban, Zhang, Bond, and Richman}]{marzban_iv}
Marzban, C., Zhang, Y., Bond, N., and Richman, M.: A Practical Introduction to
  Regression-based Causal Inference in Meteorology (II): Unmeasured
  Confounders, submitted to Advances in Statistical Climatology, Meteorology
  and Oceanography, 2026.

\bibitem[{Massmann et~al.(2021)Massmann, Gentine, and Runge}]{massmann}
Massmann, A., Gentine, P., and Runge, J.: Causal inference for process
  understanding in Earth sciences, arXiv preprint arXiv:2102.XXXXX, 2021.

\bibitem[{Melkas et~al.(2021)Melkas, Savvides, Chandramouli, M{\"a}kel{\"a},
  Nieminen, and Mammarella}]{melkas}
Melkas, L., Savvides, R., Chandramouli, S.~H., M{\"a}kel{\"a}, J., Nieminen,
  T., and Mammarella, I.: Interactive causal structure discovery in Earth
  system sciences, Journal of Machine Learning Research, pp. 1--23, 2021.

\bibitem[{Mesinger et~al.(2006)Mesinger, DiMego, Kalnay, Mitchell, Shafran,
  Ebisuzaki, Jovi{\'c}, Woollen, Rogers, Berbery, Ek, Fan, Grumbine, Higgins,
  Li, Lin, Manikin, Parrish, and Shi}]{mesinger}
Mesinger, F., DiMego, G., Kalnay, E., Mitchell, K., Shafran, P.~C., Ebisuzaki,
  W., Jovi{\'c}, D., Woollen, J., Rogers, E., Berbery, E.~H., Ek, M.~B., Fan,
  Y., Grumbine, R., Higgins, W., Li, H., Lin, Y., Manikin, G., Parrish, D., and
  Shi, W.: North American Regional Reanalysis, Bulletin of the American
  Meteorological Society, 87, 343--360, \doi{10.1175/BAMS-87-3-343}, data
  available at
  \url{https://www.ncei.noaa.gov/products/weather-climate-models/north-american-regional},
  2006.

\bibitem[{Morgan and Winship(2007)}]{morgan_book}
Morgan, S.~L. and Winship, C.: Counterfactuals and Causal Inference: Methods
  and Principles for Social Research, Cambridge University Press, 2007.

\bibitem[{Mucesh et~al.(2024)Mucesh, Hartley, Gilligan-Lee, and Lahav}]{galaxy}
Mucesh, S., Hartley, W.~G., Gilligan-Lee, C.~M., and Lahav, O.: Nature versus
  nurture in galaxy formation: the effect of environment on star formation with
  causal machine learning, \urlprefix\url{https://arxiv.org/abs/2412.02439},
  2024.

\bibitem[{Neyman(1935)}]{neyman}
Neyman, J.: Statistical problems in agricultural experimentation (with
  discussion), Supplement to the Journal of the Royal Statistical Society, 2,
  107--180, 1935.

\bibitem[{Noah(2021)}]{noah}
Noah: Why do we do matching for causal inference vs regressing on confounders?,
  \url{https://stats.stackexchange.com/q/544958}, cross Validated, 2021.

\bibitem[{Nowack et~al.(2020)Nowack, Runge, Eyring, and Haigh}]{nowack}
Nowack, P., Runge, J., Eyring, V., and Haigh, J.~D.: Causal networks for
  climate model evaluation and constrained projections, Nature Communications,
  \doi{10.1038/s41467-020-15195-y}, 2020.

\bibitem[{Oganisian(2026)}]{oganisian}
Oganisian, A.: Stress-Testing Assumptions: A Guide to Bayesian Sensitivity
  Analyses in Causal Inference,
  \urlprefix\url{https://arxiv.org/abs/2602.23640}, 2026.

\bibitem[{Ombadi et~al.(2020)Ombadi, Nguyen, Sorooshian, and Hsu}]{ombadi}
Ombadi, M., Nguyen, P., Sorooshian, S., and Hsu, K.-l.: Evaluation of Methods
  for Causal Discovery in Hydrometeorological Systems, Water Resources
  Research, 56, e2020WR027\,251, \doi{https://doi.org/10.1029/2020WR027251},
  e2020WR027251 2020WR027251, 2020.

\bibitem[{Papadogeorgou et~al.(2022)Papadogeorgou, Imai, Lyall, and
  Li}]{papadogeorgou}
Papadogeorgou, G., Imai, K., Lyall, J., and Li, F.: Temporal data: Estimating
  the effects of airstrikes on insurgent violence in Iraq, Journal of the Royal
  Statistical Society: Series B (Statistical Methodology), 84, 1969--1999,
  \doi{10.1111/rssb.12548}, 2022.

\bibitem[{Pearl(2009)}]{pearl2009}
Pearl, J.: Causal inference in statistics: An overview, Statistics Surveys, 3,
  96--146, \doi{10.1214/09-SS057}, 2009.

\bibitem[{Pearl(2010{\natexlab{a}})}]{pearl2010}
Pearl, J.: The foundations of causal inference, Sociological Methodology, 40,
  75--149, 2010{\natexlab{a}}.

\bibitem[{Pearl(2010{\natexlab{b}})}]{pearl_consistency}
Pearl, J.: On the Consistency Rule in Causal Inference: Axiom, Deﬁnition,
  Assumption, or Theorem?, Epidemiology, 21(6), 872--875, 2010{\natexlab{b}}.

\bibitem[{Pearl et~al.(2019)Pearl, Glymour, and Jewell}]{pearl2019}
Pearl, J., Glymour, M., and Jewell, N.~P.: Causal Inference in Statistics, John
  Wiley \& Sons, Ltd., 2019.

\bibitem[{{R Core Team}(2021)}]{Rcore}
{R Core Team}: R: A Language and Environment for Statistical Computing, R
  Foundation for Statistical Computing, Vienna, Austria,
  \urlprefix\url{https://www.R-project.org/}, 2021.

\bibitem[{Reich et~al.(2021)Reich, Yang, Guan, Giffin, Miller, and
  Rappold}]{reich}
Reich, B.~J., Yang, S., Guan, Y., Giffin, A.~B., Miller, M.~J., and Rappold,
  A.: A review of spatial causal inference methods for environmental and
  epidemiological applications, International Statistical Review, 89, 605--634,
  \doi{10.1111/insr.12452}, 2021.

\bibitem[{Richardson and Robins(2013)}]{richardson_robins}
Richardson, T.~S. and Robins, J.~M.: Single World Intervention Graphs (SWIGs):
  A unification of the counterfactual and graphical approaches to causality,
  Tech. Rep. 128(30), Center for Statistics and the Social Sciences, University
  of Washington, 2013.

\bibitem[{Richardson et~al.(2011)Richardson, Evans, and
  Robins}]{richardson_etal}
Richardson, T.~S., Evans, R.~J., and Robins, J.~M.: Transparent
  Parametrizations of Models for Potential Outcomes, in: Bayesian Statistics 9,
  Oxford University Press, ISBN 9780199694587,
  \doi{10.1093/acprof:oso/9780199694587.003.0019}, 2011.

\bibitem[{Robins(1986{\natexlab{a}})}]{robins}
Robins, J.: A new approach to causal inference in mortality studies with a
  sustained exposure period---application to control of the healthy worker
  survivor effect, Mathematical Modelling, 7, 1393--1512, 1986{\natexlab{a}}.

\bibitem[{Robins(1986{\natexlab{b}})}]{robins1986}
Robins, J.~M.: A new approach to causal inference in mortality studies with a
  sustained exposure period—application to control of the healthy worker
  survivor effect, Mathematical Modelling, 7, 1393--1512,
  \urlprefix\url{https://api.semanticscholar.org/CorpusID:72854377},
  1986{\natexlab{b}}.

\bibitem[{Rosenbaum and Rubin(1983)}]{rosenbaum_rubin}
Rosenbaum, P.~R. and Rubin, D.~B.: The central role of the propensity score in
  observational studies for causal effects, Biometrika, 70, 41--55, 1983.

\bibitem[{Rubin(1973{\natexlab{a}})}]{rubin_a}
Rubin, D.~B.: Matching to remove bias in observational studies, Biometrics, 29,
  159--183, a, 1973{\natexlab{a}}.

\bibitem[{Rubin(1973{\natexlab{b}})}]{rubin_b}
Rubin, D.~B.: The use of matched sampling and regression adjustment to remove
  bias in observational studies, Biometrics, 29, 185--203, b,
  1973{\natexlab{b}}.

\bibitem[{Rubin(2007)}]{rubin2007}
Rubin, D.~B.: The design versus the analysis of observational studies for
  causal effects: Parallels with the design of randomized trials, Statistics in
  Medicine, 26, 20--36, 2007.

\bibitem[{Stuart(2010)}]{stuart}
Stuart, E.~A.: Matching methods for causal inference: A review and a look
  forward, Statistical Science, 25, 1--21, 2010.

\bibitem[{Sugihara et~al.(2012)Sugihara, May, Ye, Hsieh, Deyle, Fogarty, and
  Munch}]{sugihara}
Sugihara, G., May, R., Ye, H., Hsieh, C.-H., Deyle, E., Fogarty, M., and Munch,
  S.: Detecting causality in complex ecosystems, Science, 338, 496--500, 2012.

\bibitem[{Sävje(2023)}]{savje}
Sävje, F.: Causal inference with misspecified exposure mappings: separating
  definitions and assumptions, Biometrika, 111, 1--15,
  \doi{10.1093/biomet/asad019}, 2023.

\bibitem[{Tsonis et~al.(2018)Tsonis, Deyle, Ye, and Sugihara}]{tsonis}
Tsonis, A.~A., Deyle, E.~R., Ye, H., and Sugihara, G.: Convergent Cross
  Mapping: Theory and an Example, in: Advances in Nonlinear Geosciences, edited
  by Tsonis, A.~A., pp. 587--600, Springer International Publishing, Cham, ISBN
  978-3-319-58895-7, \doi{10.1007/978-3-319-58895-7_27}, 2018.

\bibitem[{VanderWeele(2009)}]{vanderweele_consistency}
VanderWeele, T.~J.: Concerning the consistency assumption in causal inference,
  Epidemiology, 20, 880--883, 2009.

\bibitem[{VanderWeele(2016)}]{mediation}
VanderWeele, T.~J.: Mediation Analysis: A Practitioner's Guide., Annual review
  of public health, 37, 17--32,
  \urlprefix\url{https://api.semanticscholar.org/CorpusID:42762783}, 2016.

\bibitem[{VanderWeele and Shpitser(2013)}]{vanderWeele}
VanderWeele, T.~J. and Shpitser, I.: On the definition of a confounder, The
  Annals of Statistics, 41, 196--220, \doi{10.1214/12-AOS1058}, 2013.

\bibitem[{Vansteelandt et~al.(2010)Vansteelandt, Bekaert, and
  Claeskens}]{vansteelandt}
Vansteelandt, S., Bekaert, M., and Claeskens, G.: On model selection and model
  misspecification in causal inference, Statistical Methods in Medical
  Research, 21, 7--30, 2010.

\bibitem[{Xiao and Wu(2025)}]{xiao_wu}
Xiao, Z. and Wu, P.: Causal Inference in Panel Data with a Continuous
  Treatment, \urlprefix\url{https://arxiv.org/abs/2506.23226}, 2025.

\bibitem[{Zanga et~al.(2022)Zanga, Ozkirimli, and Stella}]{zanga}
Zanga, A., Ozkirimli, E., and Stella, F.: A survey on causal discovery: Theory
  and practice, International Journal of Approximate Reasoning, 151, 101--129,
  2022.

\bibitem[{Zeder and Fischer(2023)}]{zeder}
Zeder, J. and Fischer, E.~M.: Quantifying the statistical dependence of
  mid-latitude heatwave intensity and likelihood on prevalent physical drivers
  and climate change, Advances in Statistical Climatology, Meteorology and
  Oceanography, 9, 83--102, \doi{10.5194/ascmo-9-83-2023}, 2023.

\bibitem[{Zeng and Wang(2022)}]{zeng_wang}
Zeng, J. and Wang, R.: A survey of causal inference frameworks, arXiv preprint
  arXiv:2209.00869, \doi{10.48550/arXiv.2209.00869}, 2022.

\end{thebibliography}

\end{document}